\documentclass[twocolumn,prc,superscriptaddress,amsmath,amssymb,nofootinbib]{revtex4}

\usepackage{graphicx}
\usepackage{dcolumn}
\usepackage{bm}
\usepackage{xcolor, color, soul}
\usepackage{mathptmx}
\usepackage{multirow}
\usepackage{longtable}
\usepackage{mwe}
\usepackage{footnotehyper}
\usepackage[
            pdfstartview=FitH,
            CJKbookmarks=true,
            bookmarksnumbered=true,
            bookmarksopen=true,
            colorlinks=true,
            linkcolor=blue,
            anchorcolor=blue,
            citecolor=blue
            ]{hyperref}

\begin{document}

\title{Evolution of chirality from transverse wobbling in $^{135}$Pr}

\author{N. Sensharma}\email{nsensharma@anl.gov}\thanks{Present address: Physics Division, Argonne National Laboratory, Argonne, IL 60439, USA}
\affiliation{Department of Physics and Astronomy, University of Notre Dame, Notre Dame, IN 46556, USA}
\affiliation{Department of Physics and Astronomy, University of North Carolina Chapel Hill, NC 27599, USA}
\affiliation{Triangle Universities Nuclear Laboratory, Duke University, Durham, NC 27708, USA}

\author{U. Garg}\email{garg@nd.edu}
\affiliation{Department of Physics and Astronomy, University of Notre Dame, Notre Dame, IN 46556, USA}

\author{Q. B. Chen}
\affiliation{Department of Physics, East China Normal University, Shanghai 200241, China}

\author{S. Frauendorf}\email{sfrauend@nd.edu}
\affiliation{Department of Physics and Astronomy, University of Notre Dame, Notre Dame, IN 46556, USA}

\author{S. Zhu}\thanks{Deceased}
\affiliation{National Nuclear Data Center, Brookhaven National Laboratory, Upton, NY 11973, USA}

\author{J. Arroyo}
\affiliation{Department of Physics and Astronomy, University of Notre Dame, Notre Dame, IN 46556, USA}

\author{A. D. Ayangeakaa}
\affiliation{Department of Physics and Astronomy, University of North Carolina Chapel Hill, NC 27599, USA}
\affiliation{Triangle Universities Nuclear Laboratory, Duke University, Durham, NC 27708, USA}

\author{D. P. Burdette}\thanks{Present address: Physics Division, Argonne National Laboratory, Argonne, IL 60439, USA}
\affiliation{Department of Physics and Astronomy, University of Notre Dame, Notre Dame, IN 46556, USA}

\author{M. P. Carpenter}
\author{P. Copp}\thanks{Los Alamos National Laboratory, Los Alamos,
New Mexico 87545, USA}
\affiliation{Physics Division, Argonne National Laboratory, Argonne, IL 60439, USA}

\author{J. L. Cozzi}\thanks{Present address: Department of Radiology, University of Chicago, Chicago, IL 60637, USA}
\affiliation{Department of Physics and Astronomy, University of Notre Dame, Notre Dame, IN 46556, USA}

\author{S. S. Ghugre}
\affiliation{UGC-DAE Consortium for Scientific Research, Kolkata 700 064, India}

\author{D. J. Hartley}
\affiliation{Department of Physics, United States Naval Academy, Annapolis, MD 21402, USA}

\author{K. B. Howard}\thanks{Present address: Department of Physical Science, Anne Arundel Community College, Arnold, MD 21012, USA}
\affiliation{Department of Physics and Astronomy, University of Notre Dame, Notre Dame, IN 46556, USA}

\author{R. V. F. Janssens}
\affiliation{Department of Physics and Astronomy, University of North Carolina Chapel Hill, NC 27599, USA}
\affiliation{Triangle Universities Nuclear Laboratory, Duke University, Durham, NC 27708, USA}

\author{F. G. Kondev}
\author{T. Lauritsen}
\author{J. Li}
\affiliation{Physics Division, Argonne National Laboratory, Argonne, IL 60439, USA}

\author{R. Palit}
\affiliation{Department of Nuclear and Atomic Physics, Tata Institute of Fundamental Research, Mumbai 400 005, India}

\author{A. Saracino}
\affiliation{Department of Physics and Astronomy, University of North Carolina Chapel Hill, NC 27599, USA}
\affiliation{Triangle Universities Nuclear Laboratory, Duke University, Durham, NC 27708, USA}

\author{D. Seweryniak}
\affiliation{Physics Division, Argonne National Laboratory, Argonne, IL 60439, USA}

\author{S. Weyhmiller}\thanks{Present address: Physics Department, Yale University, New Haven, CT 06511, USA}
\affiliation{Department of Physics and Astronomy, University of Notre Dame, Notre Dame, IN 46556, USA}

\author{J. Wu}
\affiliation{Physics Division, Argonne National Laboratory, Argonne, IL 60439, USA}

\begin{abstract}

Chirality is a distinct signature that characterizes triaxial shapes in nuclei. We report the first observation of chirality in the nucleus $^{135}$Pr using a high-statistics Gammasphere experiment with the $^{123}$Sb($^{16}$O,4n)$^{135}$Pr reaction. Two chiral-partner bands with the configuration $\pi(1h_{11/2})^1\otimes \nu(1h_{11/2})^{-2}$ have been identified in this nucleus. Angular distribution analyses of the $\Delta I = 1$ transitions connecting the two bands reveal a dominant dipole character, and quasiparticle triaxial rotor model calculations show good agreement with the data. Since the simultaneous observation of chirality and transverse wobbling in $^{135}$Pr relies critically on these angular distribution results, we also address and refute the experimental and theoretical criticisms raised in a recent work by Lv et al., presenting additional evidence that further strengthens our interpretation. This marks the first observation of both hallmarks of triaxiality—chirality and wobbling—in the same nucleus.

\end{abstract}

\date{\today}

\pacs{}

\maketitle

\section{Introduction}

\begin{figure*}[ht]
\centering
\includegraphics[width=\textwidth]{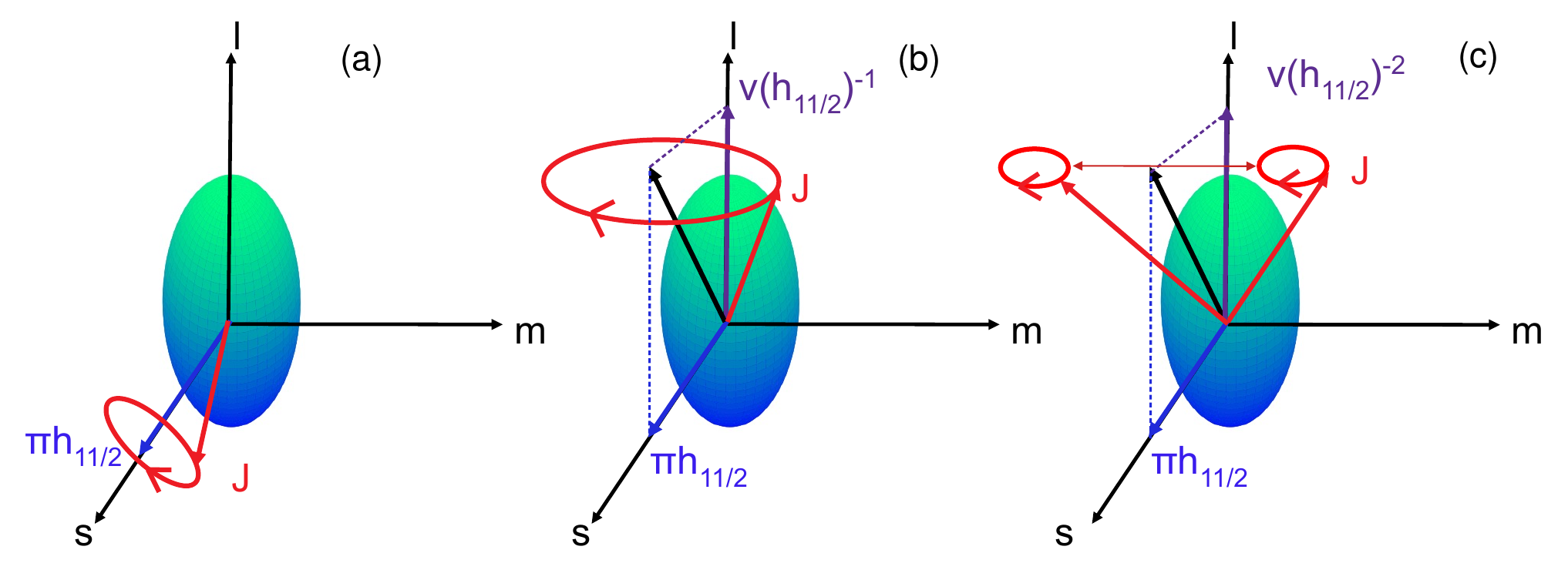}
\caption{\label{f:schematic}(Color online) Angular momentum geometry for (a) transverse wobbling (TW), (b) transverse chiral vibration (TCV), and (c) chiral rotation (CR) mode in the body fixed frame, where $l$, $m$, and $s$ correspond to the long, medium, and short axis, respectively, and $\bm{J}$ is the total angular momentum vector. The TW orbit is centered around the $s$-axis while the TCV orbit around the black axis in the $l$-$s$ plane is the total quasiparticle angular momentum. In the CR mode $\bm{J}$ is localized out of the three principal planes and oscillates between the equivalent octants of opposite chirality of the principal axes (only two are shown). The quasineutron and quasiproton angular momenta are shown as the violet and blue arrows aligned with the $l$- and $s$- axes, respectively. With increasing $J$, the quasiparticle arrows start following the motion of $\bm{J}$.} 
\end{figure*}

Nuclear wobbling motion and chiral rotation are two distinct phenomena that serve as key experimental signatures of nuclear triaxiality. In even-even triaxial nuclei, Bohr and Mottelson~\cite{bohr} conceptualized wobbling motion as the oscillation of one of the principal axes of a triaxial rotor about the space-fixed angular momentum vector. Experimentally, this rare mode has been established in the even-even nuclei: $^{112}$Ru~\cite{Hamilton2010NPA, Frauendorf2024book}, $^{114}$Pd~\cite{Y.X.Luo2013proceeding}, $^{130}$Ba~\cite{Petrache2019PLB, 130Ba}, $^{136}$Nd~\cite{136Nd}, and $^{82}$Kr~\cite{Rajbanshi2025PRC}. On the other hand, wobbling motion has been observed in several odd-$A$ nuclei across different regions of the nuclear chart, including $^{161, 163, 165, 167}$Lu~\cite{161Lu,163Lu,165Lu,167Lu}, $^{167}$Ta~\cite{167Ta}, and $^{151}$Eu~\cite{Mukherjee2023PRC} in the $A \sim 160$ region, $^{135}$Pr~\cite{135Pr,two-phonon}, $^{133}$La~\cite{Biswas2019EPJA}, $^{129, 133}$Ba~\cite{Chakraborty2024PRC, Devi2021PLB}, $^{125, 127}$Xe~\cite{Prajapati2024PRC, Chakraborty2020PLB}, and $^{139}$Pm~\cite{Rajbanshi2024PRC} in the $A \sim 130$  region, $^{183,187}$Au~\cite{nandi,187Au} in the $A \sim 190$ region, and $^{105}$Pd \cite{105Pd} in the $A \sim 100$ region. In all these odd $A$ cases, the odd particle/hole coupled to the even-even triaxial core stabilizes the triaxial nature of the nucleus at high spins, thereby, facilitating experimental observation. Figure~\ref{f:schematic} (a) schematically illustrates these transverse wobbling (TW) structures.  All observed cases exhibit rotational bands associated with a wobbling phonon number ($n_\omega$ = 0, 1, 2, $\ldots$), with successive wobbling bands connected via $\Delta I$ = 1, $E2$ transitions.  

The other signature of triaxiality, nuclear chiral rotation, appears when the axis of rotation does not lie in any of the three principal planes of the nucleus~\cite{chiral, Frauendorf2001RMP, J.Meng2010JPG, J.Meng2016PS}. This arrangement occurs when there is a finite angular momentum component along all three principal axes of a triaxial rotor. Chirality is experimentally manifested as two nearly identical $\Delta I = 1$ sequences with the same spin, parity and close excitation energies. A number of nuclei have been found to exhibit chirality. Some examples include nuclei in the regions $A \sim 80$ ($^{80}$Br~\cite{wang}, $^{78}$Br~\cite{C.Liu2016PRL}, and $^{74}$Br~\cite{R.J.Guo2024PRL}), $A \sim 100$ ($^{104}$Rh~\cite{vaman, Krako2024PLB}, $^{103}$Rh~\cite{Kuti2014PRL}, and $^{102}$Rh~\cite{Tonev2014PRL}), $A \sim 130$ ($^{135}$Nd~\cite{zhu,Mukhopadhyay2007PRL},$^{134}$Pr~\cite{Starosta2001PRL, Tonev2006PRL, Tonev2007PRC,timar2011}, $^{128}$Cs~\cite{Grodner2006PRL, Grodner2018PRL}, and $^{133}$Ce~\cite{ayangeakaa}), and $A \sim 190$ ($^{188}$Ir~\cite{balabanski}). A comprehensive compilation can be found in Ref.~\cite{B.W.Xiong2019ADNDT}. In all cases, two $\Delta I=1$ bands with close excitation energies were reported. The small energy difference between the two bands were attributed to tunneling between the left- and right-handed configurations in the case of static chirality or chiral rotation (CR)~\cite{chiral}, or to an oscillation between the two configurations in the case of chiral vibrations (CV)~\cite{Starosta2001PRL}. The authors of Ref.~\cite{CF2025} analyzed the chiral modes in detail and classified the CV
as transverse CV (TCV) or longitudinal CV (LCV) depending on, respectively, whether the vibration plane is perpendicular to the medium ($m$) axis as in 
Fig.~\ref{f:schematic} (b) or contains it.

\begin{figure*}[ht]
\centering
\includegraphics[width=\textwidth]{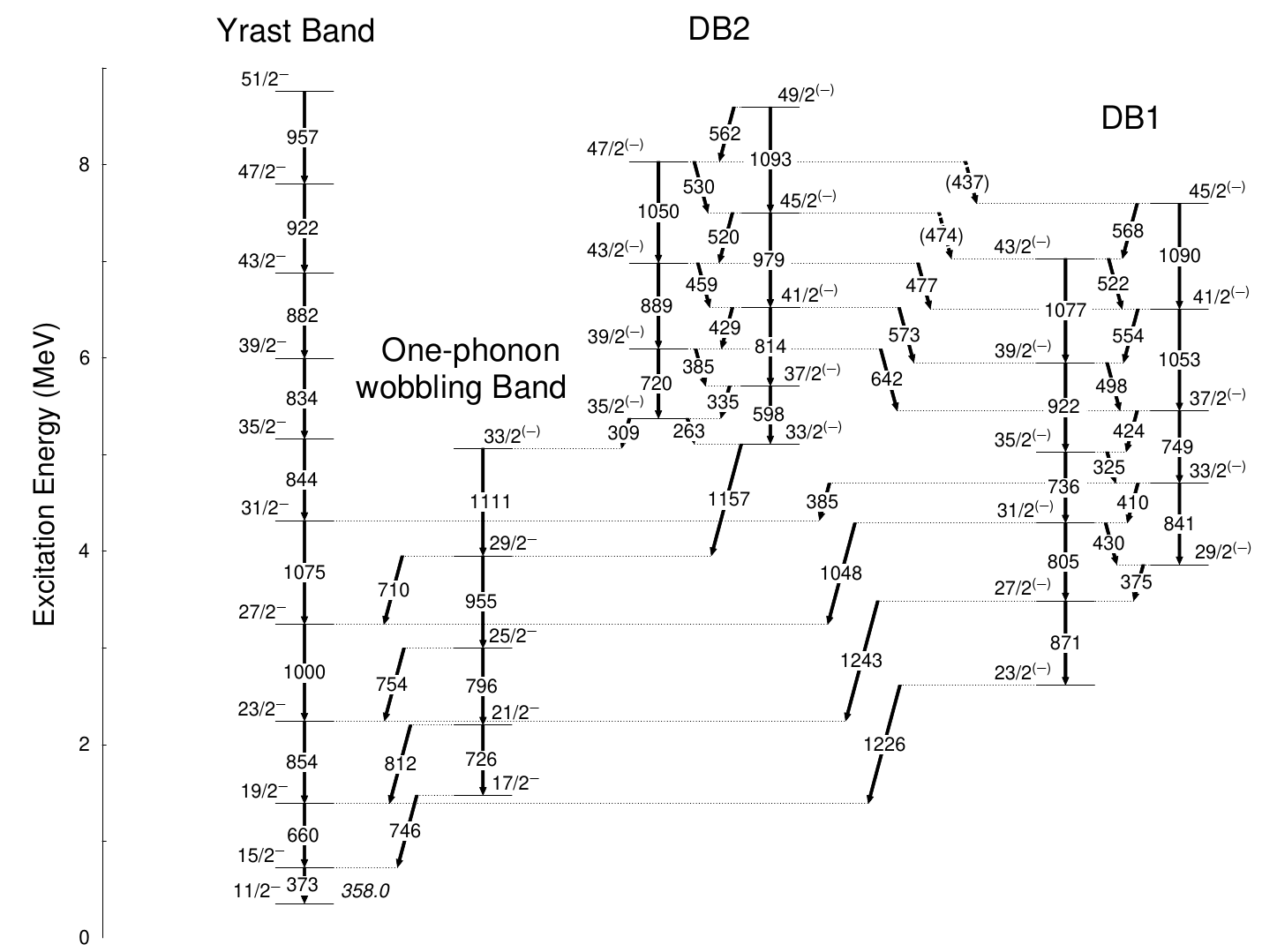}
\caption{\label{f:level_scheme} Partial level scheme of $^{135}$Pr developed in the present work. The lowest level shown is an 11/2$^{-}$ isomeric level with $E_{x}$ = 358.0 keV. The five connecting transitions between dipole bands DB2 and DB1 have been newly identified in this work. The tentative $\gamma$-ray transitions are given as dotted lines.} 
\end{figure*}

This paper presents a detailed investigation of $^{135}$Pr, a nucleus with low deformation $\epsilon$ $\sim$ 0.16~\cite{tw_frauendorf} and a triaxial core coupled to an odd-$h_{11/2}$ proton particle. Ref.~\cite{135Pr} identified $^{135}$Pr as the first wobbling nucleus within the $A \sim 130$ region, demonstrating a decreasing trend of wobbling energy with increasing angular momentum. This observation was explained within the quasiparticle triaxial rotor (QTR) model as a consequence of the alignment of the odd-$h_{11/2}$ proton particle along the short (s) axis of the rotor. This phenomenon of the odd particle aligning perpendicular to the axis with the maximum moment of inertia (i.e., the $m$ axis) has, hence, been termed as ``transverse wobbling". Following the identification of the one-phonon ($n_\omega$ = 1) transverse wobbling band in $^{135}$Pr, Ref.~\cite{two-phonon} reported the second-phonon ($n_\omega$ = 2) wobbling band, decaying to the $n_\omega$ = 1 band via four $\Delta I$ = 1, $E2$ transitions.

In addition to the $n_\omega$ = 1 transverse wobbling band, Ref.~\cite{135Pr} identified a dipole band building on top of the wobbling band. Using the tilted axis cranking and QTR models, this band was explained as a mutation of transverse wobbling into a three-quasiparticle band of the magnetic rotation type. Further extension to the level scheme of $^{135}$Pr was reported in Ref.~\cite{James_phd} and another dipole band originating at a much lower excitation energy of $E_x$ = 2616 keV with strong in-band $\Delta I$ = 1 transitions and relatively weak in-band $\Delta I$ = 2 transitions was established. 

In this paper, we have further explored the two previously reported dipole bands and identified five $\Delta I$ = 1 transitions that connect them. Based on the close excitation energies of the two bands and by determining the transition-probability ratios for the $I\rightarrow I-1$ transitions between them, the present work, building on the results of Refs.~\cite{135Pr,James_phd} and \cite{two-phonon}, brings forth new evidence for the existence of chiral geometry in the $^{135}$Pr nucleus. 

\begin{figure}
    \centering
   \includegraphics[width=\columnwidth]{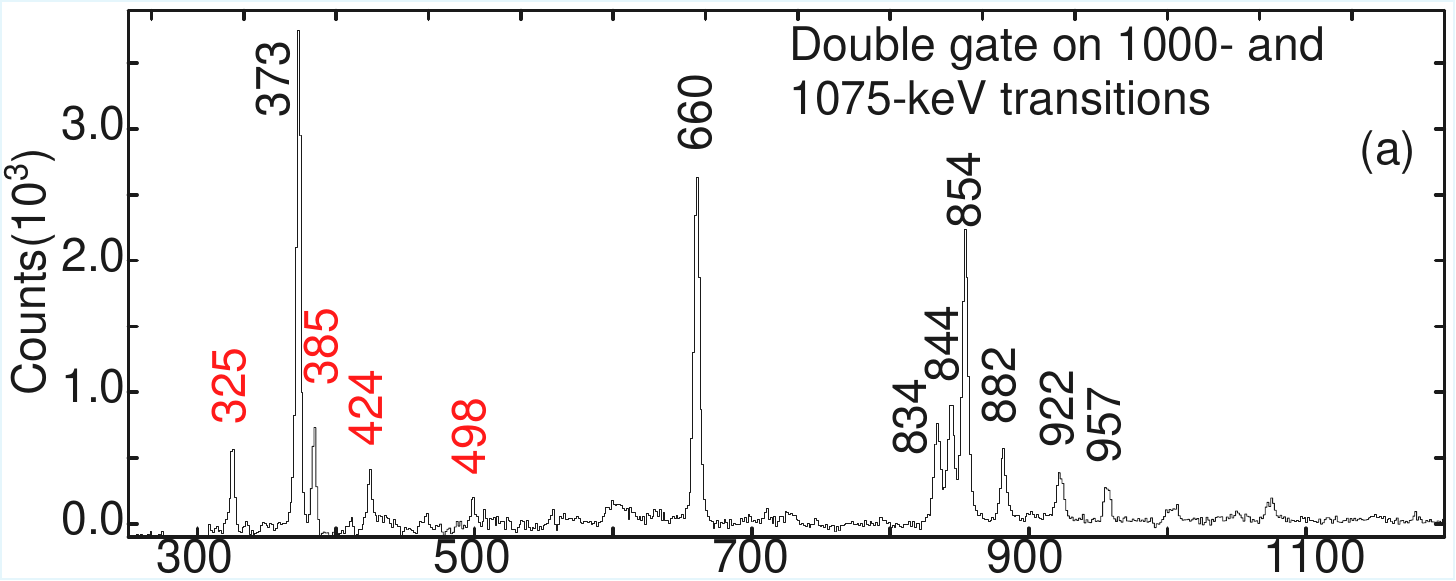}
   \includegraphics[width=\columnwidth]{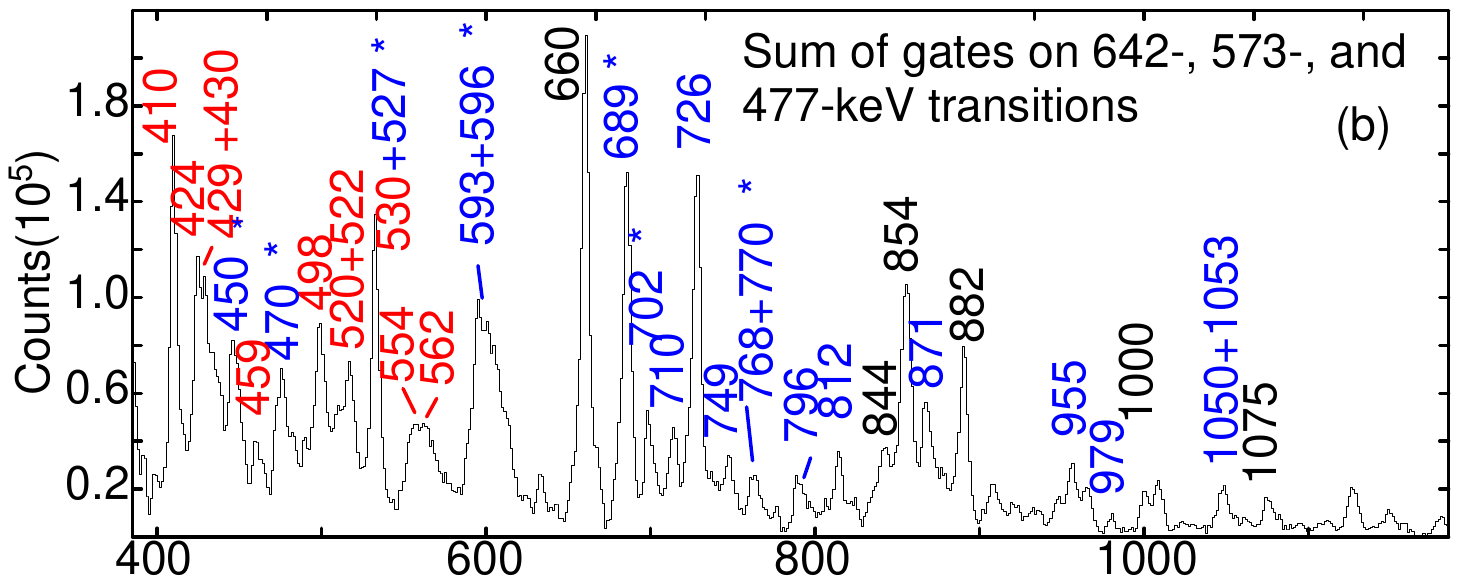}
   \includegraphics[width=\columnwidth]{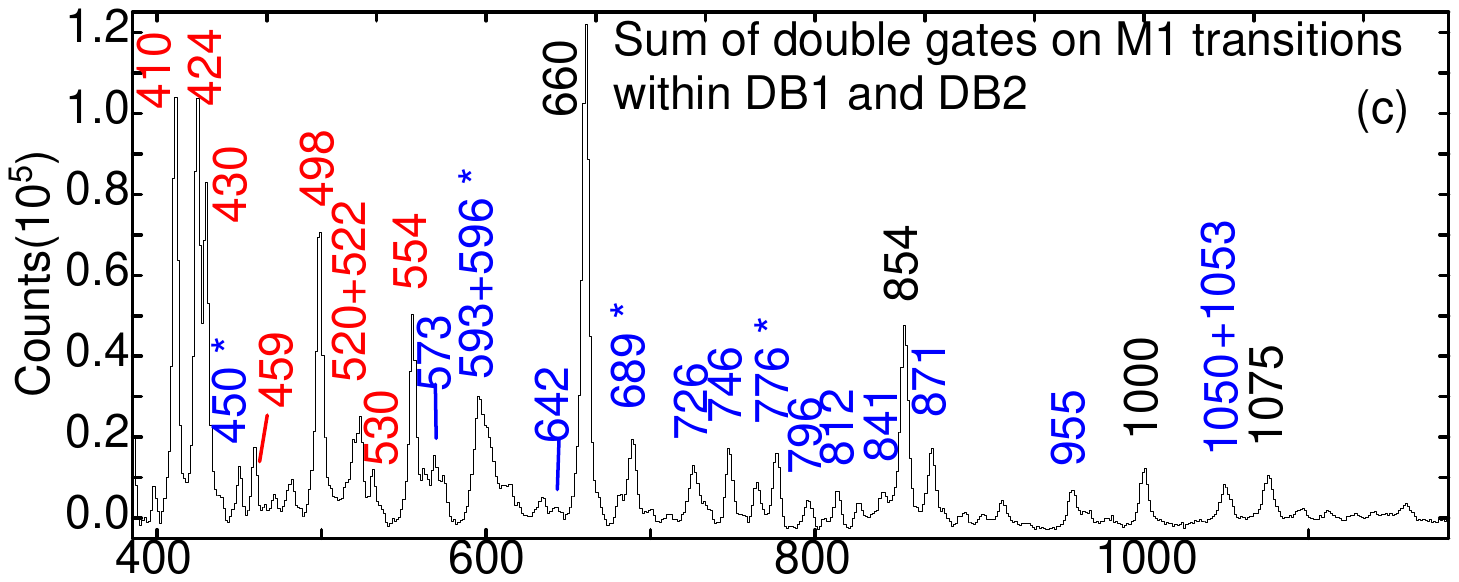}
   \includegraphics[width=\columnwidth]{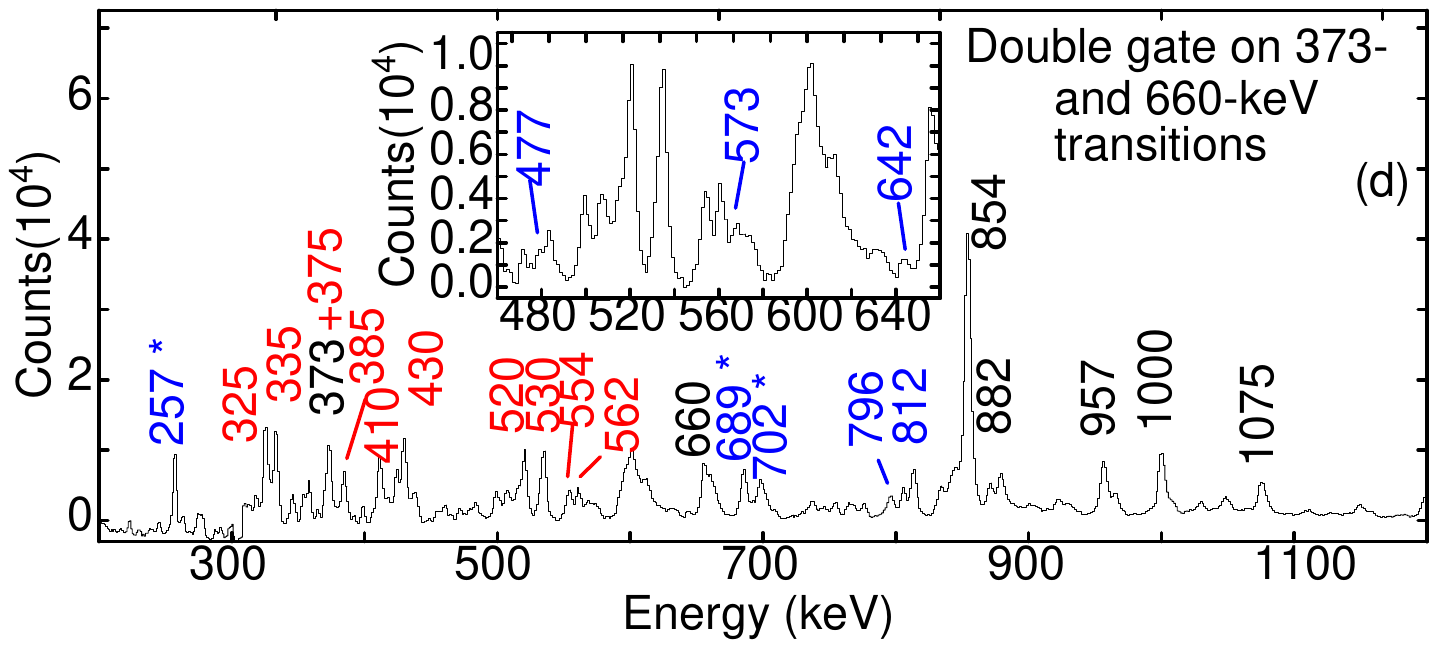}
    \caption{(Color online) The observed coincidence spectrum resulting from the (a) double gate on the yrast in-band 1000- and 1075-keV transitions, (b) sum of gates on E$_\gamma$ = 642, 573, and 477 keV, (c) sum of all possible double gates on the $\Delta I$ = 1, M1 transitions within DB1 and DB2, and (d) double gate on 373- and 660-keV transitions. An inset is included in the bottom panel to magnify and clearly display the three lowest DB2 $\to$ DB1 transitions. The coincident $\gamma$-ray energies are marked against the respective energy peaks. The peaks marked in red correspond to the $\Delta$I = 1 in-band transitions of DB1 and DB2. Yrast in-band transitions are marked in black. All other transitions arising from the deexcitation of $^{135}$Pr are marked in blue and an (*) is marked on the transitions that were observed, but are not displayed, in the level scheme of Fig.~\ref{f:level_scheme}.}
    \label{fig:spectra1}
\end{figure}

\begin{figure}[ht]
    \centering
   \includegraphics[width=\columnwidth]{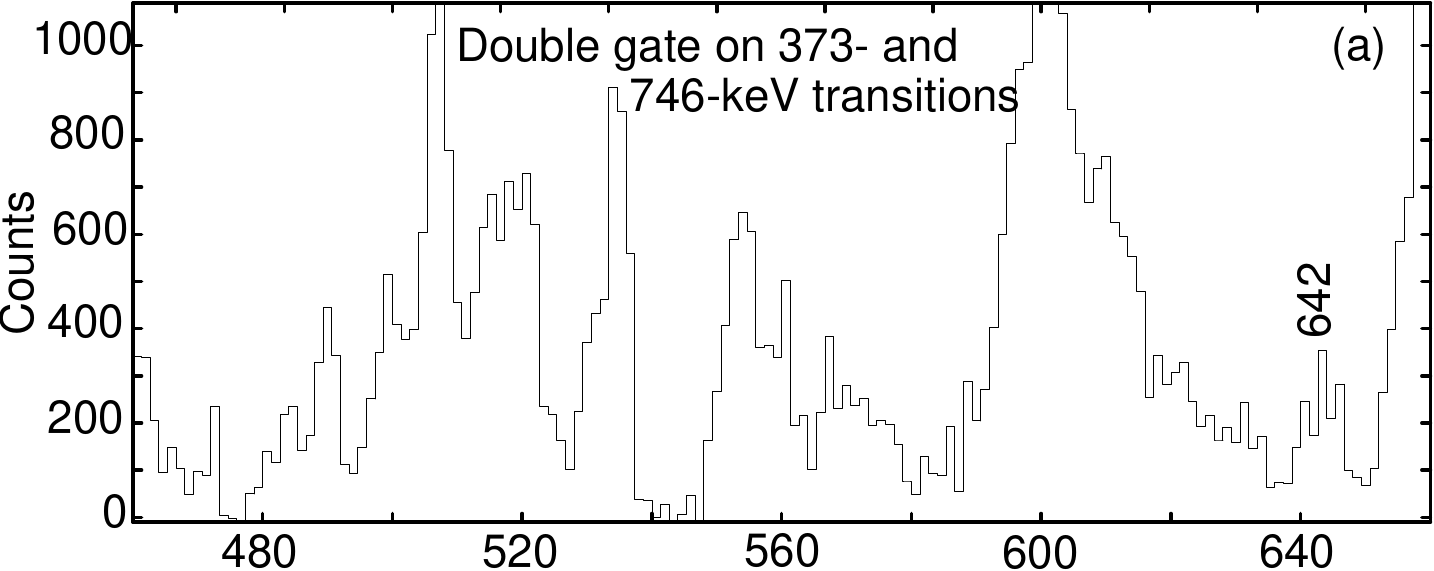}
   \includegraphics[width=\columnwidth]{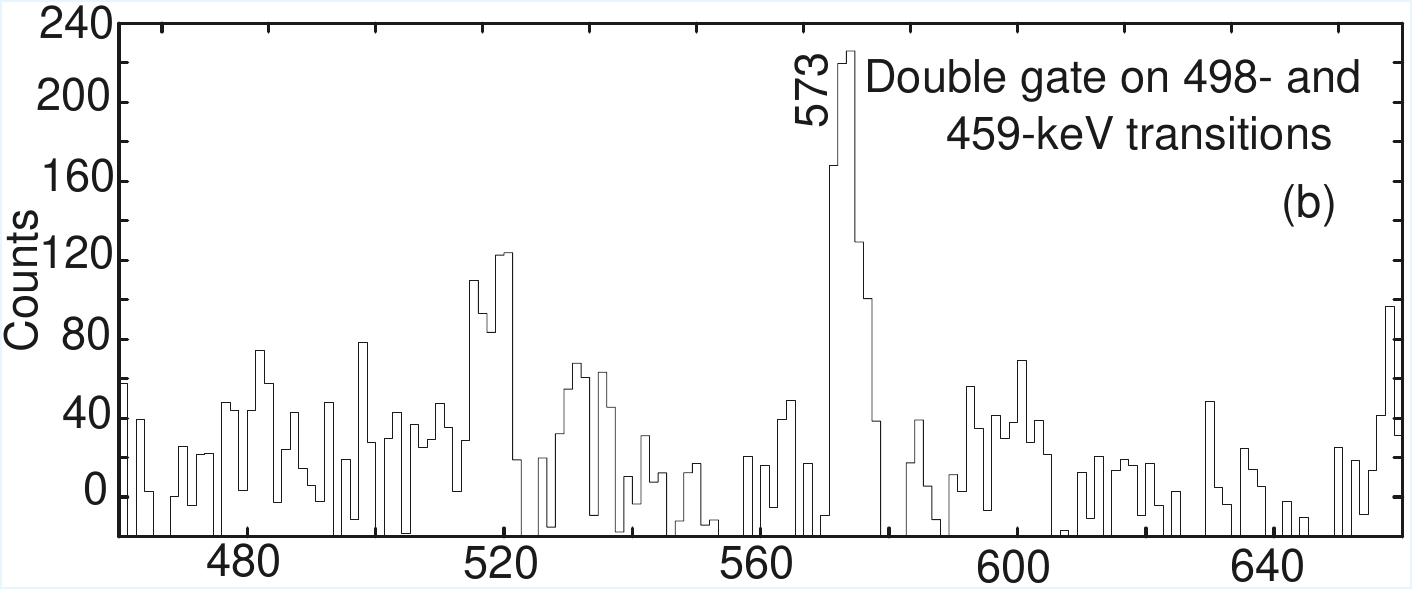}
   \includegraphics[width=\columnwidth]{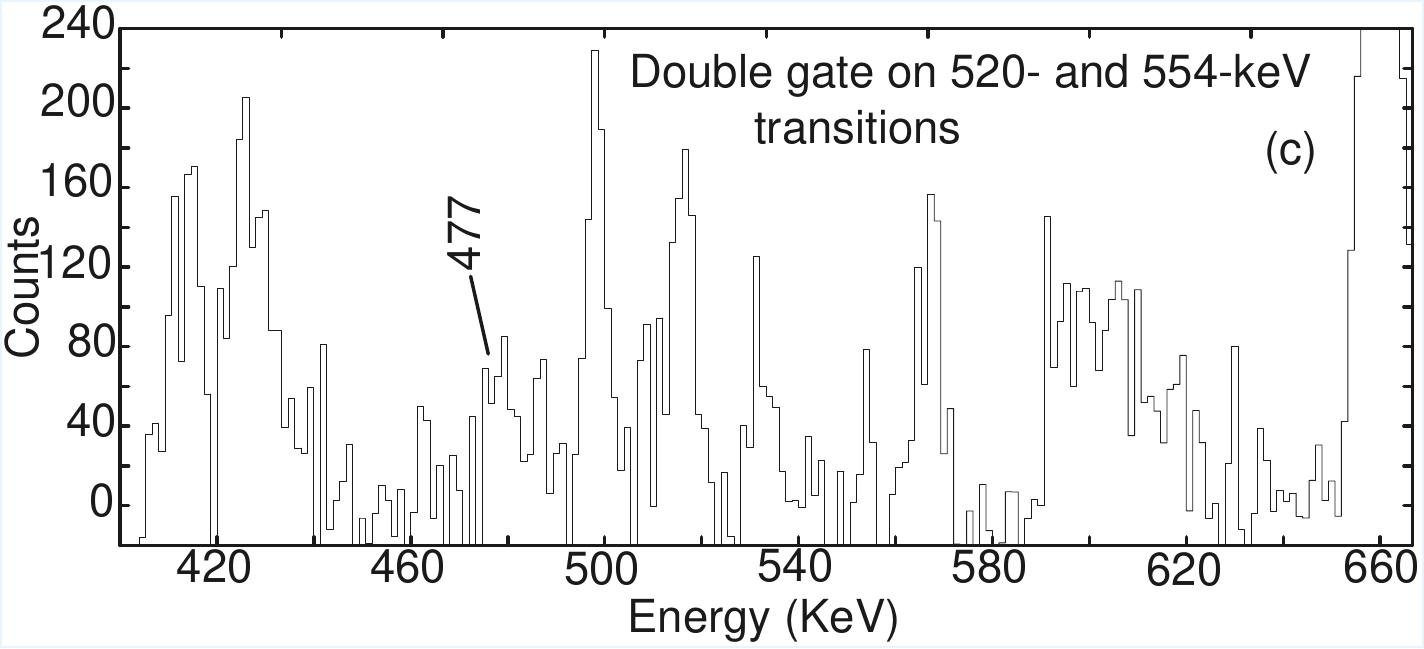}
    \caption{The observed coincidence spectrum resulting from a double gate on (a) 373- and 746-keV transitions, (b) 498- and 459-keV transitions, and (c) 520- and 554-keV transitions. The spectra are zoomed in to present the DB2 $\to$ DB1 transitions. All the other peaks observed in these spectra have been identified and placed in the level scheme.}
    \label{fig:spectra2}
\end{figure}

\section{Experiment}
This experiment is part of a series of experiments \cite{135Pr,two-phonon} performed using the Gammasphere array at the Argonne National Laboratory to investigate the structure of the $^{135}$Pr nucleus. The details of the experiment are similar to those described in Refs.~\cite{135Pr,two-phonon}. An 80-MeV $^{16}$O beam impinged on an enriched $^{123}$Sb target to populate the levels of interest in $^{135}$Pr. The target was composed of a 697~$\mu$g/cm$^{2}$-thick foil of $^{123}$Sb with a 15~$\mu$g/cm$^{2}$ front layer of aluminum. A total of 63 Compton-suppressed High Purity Germanium detectors of the Gammasphere array were available and data were acquired in the triple-coincidence mode. Energy and efficiency calibrations were performed using a standard $^{152}$Eu radioactive source. To increase our statistics, data from the present run were added to the data obtained from the run described in Ref.~\cite{two-phonon}. The combined total of three- and higher-fold $\gamma$-ray coincidence events was 2.5 $\times$ 10$^{10}$. The \texttt{RADWARE} suite of codes~\cite{radford} was utilized to analyze the combined data by sorting them into $\gamma$-$\gamma$ coincidence matrices and $\gamma$-$\gamma$-$\gamma$ coincidence cubes. A partial level scheme for $^{135}$Pr relevant for the focus of this work is presented in Fig.~\ref{f:level_scheme}. In addition, all level and $\gamma$ energies, initial and final spins, $\gamma$ intensities, and other parameters extracted in this work are displayed in Table. \ref{tab:1}.

The high-spin yrast states in $^{135}$Pr were firmly established by Ref.~\cite{135Pr_yrast} up to spin $67/2^-$ and extended to tentative spin $(91/2)^-$. The current study aligns with these placements, and Fig.~\ref{fig:spectra1} (a) presents the corresponding coincidence spectrum, showcasing the yrast in-band transitions up to spin $51/2^-$. The two dipole bands, as mentioned earlier, labeled DB1 and DB2 in Fig.~\ref{f:level_scheme}, were first identified in Refs.~\cite{James_phd} and \cite{135Pr}, respectively. The present work has confirmed the placement of all the $\gamma$ rays within the two bands, and has identified five new $\Delta I = 1$ transitions connecting them. Figure~\ref{fig:spectra1} (b) presents the coincidence spectrum gated on the sum of the three lowest newly identified DB2 $\to$ DB1 transitions (viz. 642.2, 572.9, and 476.9 keV). It must be noted here that this combination of sum gates shows a considerable enhancement of the 726-keV $\gamma$ ray which is not in agreement with the relative intensity assignments listed in Table \ref{tab:1}. This indicates the presence of an additional contribution to the 726-keV peak, likely arising from weakly populated decay paths in $^{135}$Pr or from a minor contribution from another reaction channel. While the precise origin of this additional intensity could not be uniquely identified, its presence does not affect the level placement or the main conclusions of the present work. Figure~\ref{fig:spectra1} (c) displays the spectrum resulting from the sum of all possible coincidence double gates on the $M1$ in-band transitions within the two dipole bands, and Fig.~\ref{fig:spectra1} (d) presents coincidence spectrum resulting from a double gate on the 373- and 660-keV transitions. These spectra confirm the existence of DB1 and DB2 bands, as well as the connecting transitions between them, within the deexcitation scheme of $^{135}$Pr. To enhance the visualization of the weak DB2 $\to$ DB1 connecting transitions, additional coincidence spectra are presented in Fig.~\ref{fig:spectra2}. The top panel (a) shows the spectrum obtained from a double gate on the 373- and 746-keV transitions, highlighting the 642-keV transition, the middle panel (b) displays the spectrum gated on the 498- and 459-keV transitions, revealing the 573-keV transition, while the bottom panel (c) displays a double-gated spectrum on the 520- and 554-keV transitions to show the 477-keV DB2 $\to$ DB1 connecting transition. 

\begin{figure*}
\centering
\includegraphics[width=\textwidth]{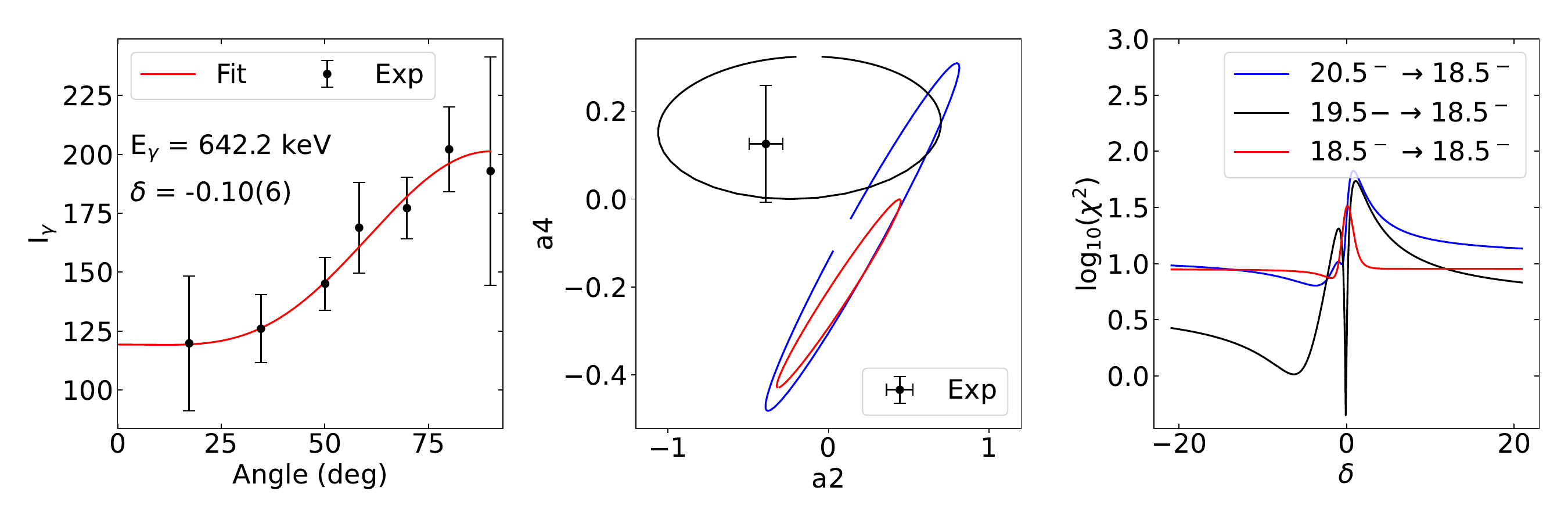}\\
\includegraphics[width=\textwidth]{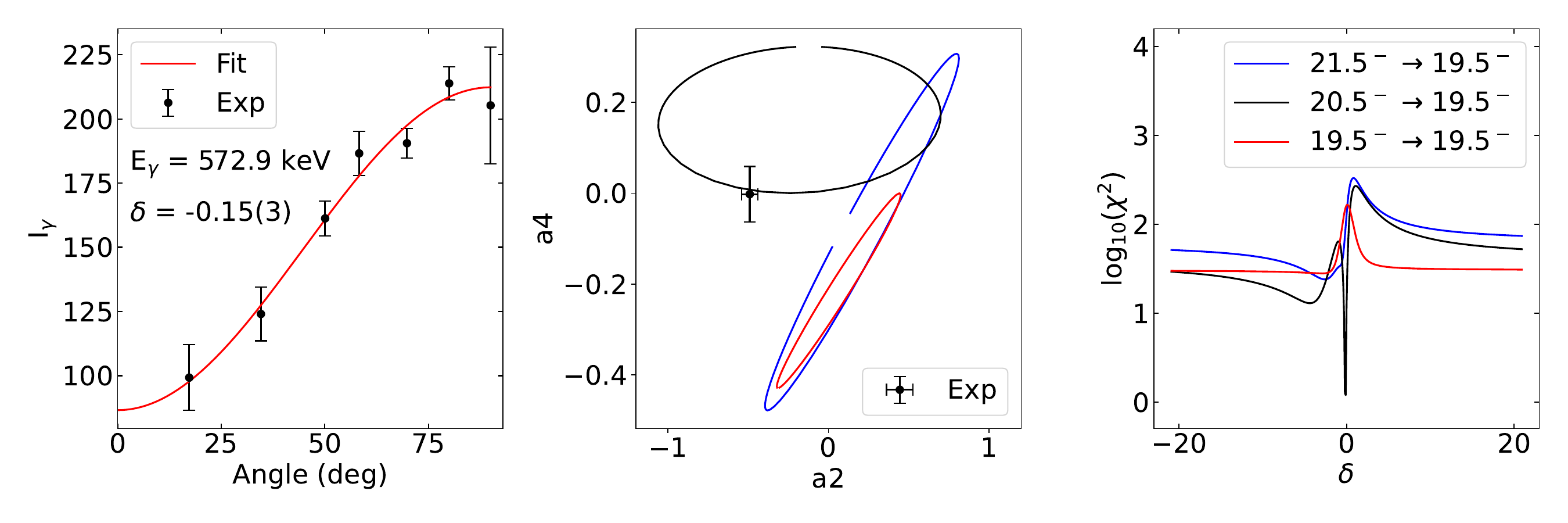}\\
\includegraphics[width=\textwidth]{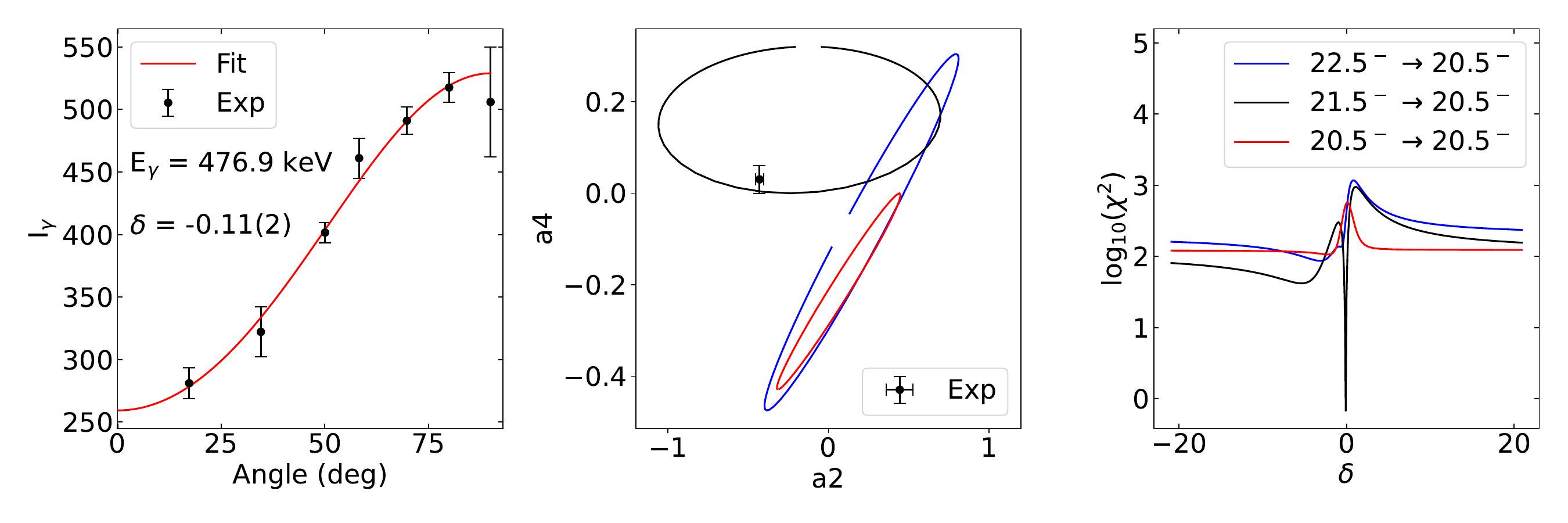}
\caption{\label{f:angdis}(Color online) Angular distribution plots for the three lowest $\Delta I = 1$ transitions connecting the DB1 and DB2 bands. (Left) The experimental points are given as black circles, and the solid red lines are fits to the angular distributions. (Center) A plot of the a$_4$ coefficient versus the a$_2$ coefficient for various initial and final spin combinations. The experimental a$_4$-a$_2$ values, obtained from the fits in the left panel, are displayed in black and align with the most probable spin sequence. (Right) The calculated $\chi^2$ values comparing theoretical and experimental angular distributions from the left panel. The minimum $\chi^2$ values in these plots correspond to the $\delta$ values extracted from the angular distribution fits. For completeness, $\chi^2$ values for other possible spin sequences are also included.} 
\end{figure*}

Angular distribution measurements were carried out for several of the transitions shown in Fig. \ref{f:level_scheme}, using the combined data set. In order to account for the different number of detectors between the two experimental runs, we combined the total $\gamma$-ray yield from both datasets and constructed a total efficiency file that reflects the detector configuration and live time for each run. The angular distributions were then corrected using this combined efficiency file on a ring-by-ring basis. This procedure ensures that the angular intensities are appropriately normalized to the effective detection efficiency at each angle, eliminating any potential skew arising from the differing detector coverage. For the three lowest $\Delta I = 1$ interconnecting transitions, the mixing ratios ($\delta$) corresponding to the lowest $\chi^2$ values were extracted. The details of the angular distribution measurements are the same as those described in Refs.~\cite{two-phonon, au-prl}. Figure~\ref{f:angdis} (left panel) displays the angular distributions for the 642.2-, 572.9-, and 476.9-keV transitions connecting the DB2 and DB1 bands, with the corresponding $\delta$ value noted on each plot. These angular distributions were measured in gated spectra - for the 642-keV transition, the angular distribution was obtained with a gate on the strong 660-keV transition in order to remove the contamination arising from it, for the 573-keV transition, the gate was applied on the 498-keV DB1 in-band transition, while for the 477-keV transition, the applied gate was at the 554-keV DB1 in-band transition. It is seen from these angular distributions that the extracted values of $\delta$ are small, implying that these transitions are primarily dipole in character. To obtain the angular distribution coefficients ($a_2$, $a_4$) and extract the mixing ratio $\delta$, a Markov Chain Monte Carlo (MCMC) technique was employed. In this approach, $a_2$ and $a_4$ are initially sampled as independent parameters across the physically allowed range using a random walker algorithm. From the resulting $a_2$ values, quadratic equations in $\delta$ are constructed and solved to obtain all physically meaningful solutions. The solution yielding the lower $\chi^2$ value is selected, and further sampling is carried out in its vicinity to refine the estimates of $a_2$, $a_4$, and $\delta$ by calculating the associated $\chi^2$ likelihood at each step. This method ensures a statistically robust extraction of angular distribution parameters even for weak transitions. Further details of the MCMC methodology can be found in Ref.~\cite{MCMCMethod}. To rule out other possible spin sequences, the center and right panels of this figure present the variation of theoretical angular distribution coefficients ($a_2$ and $a_4$) for different initial and final spin combinations, along with the corresponding $\chi^2$ values as a function of the mixing ratio. The most probable spin sequence is identified by the overlap of the experimentally determined $a_2$-$a_4$ coefficients with the theoretical predictions. Additionally, the chosen spin sequence for each $\gamma$ transition corresponds to the $\chi^2$ distribution exhibiting the lowest $\chi^2$ value. Since polarization measurements were not possible with the present experimental setup, parity assignments of all levels within the DB1 and DB2 bands have been marked as tentative in the level scheme provided in Fig.~\ref{f:level_scheme}. The angular distributions presented in this work, however, firmly assign the multipolarity of the DB2 $\to$ DB1 transitions. In addition to the DB2 $\to$ DB1 transitions, angular distributions were also performed for several other transition depopulating in the scheme of $^{135}$Pr. The $\delta$ values extracted from these distributions are provided in Table \ref{tab:1}.

\begin{figure*}
\centering
\includegraphics[width=0.8\textwidth]{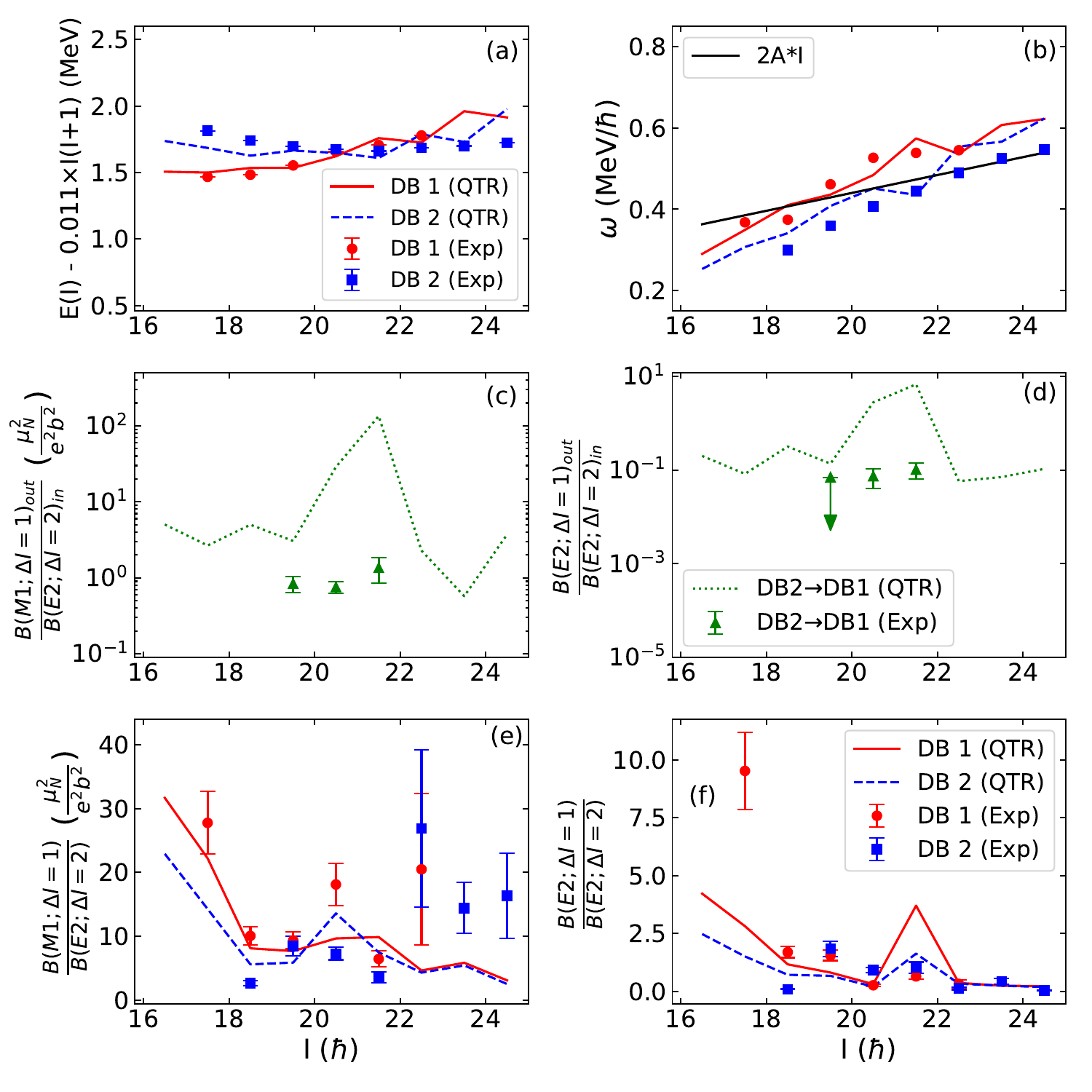}
\caption{\label{f:energy}(Color online) QTR results in comparison with the experimental data: (a) excitation energies minus a reference, $E(I) - 0.011\times I(I+1)$, (b) Rotational frequency $\hbar \omega(I-1/2)=[E(I)-E(I-2)]/2$, where the black straight line shows $ \omega=2A I$ with $A=0.011$MeV/$\hbar^2$, (c) $B(M1; \Delta I = 1)_{out}/B(E2; \Delta I = 2)_{in}$ and (d) $B(E2; \Delta I = 1)_{out}/B(E2; \Delta I = 2)_{in}$ ratios for the transitions connecting the DB2 and DB1 bands, (e) $B(M1; I \to I - 1)/B(E2; I \to I - 2)$, and (f) $B(E2; I \to I- 1)/B(E2; I \to I - 2)$ ratios for the DB1 and DB2 bands. Note that the experimental error bars in (a) and (b) are too small to be visible.} 
\end{figure*}

The experimental energies for the two dipole bands are displayed in Fig.~\ref{f:energy}(a). These DB1 and DB2 bands cross each other with a minimal energy difference of 17 keV at $I$ = $41/2$. In accordance, the rotational frequency $\omega(I)$ in Fig.~\ref{f:energy}(b) indicates that DB2 has a surplus of $\approx$ 2$\hbar$ of angular momentum at a given $\omega(I)$ value. It increases below $I$ = $41/2$ and stays approximately constant above this value. The mixing ratios extracted from angular distributions were utilized to obtain the transition-probability ratios, $B(E2,I\rightarrow I-1)/B(E2,I\rightarrow I-2)$ and $B(M1,I\rightarrow I-1)/B(E2,I\rightarrow I-2)$, which are listed in the last two columns of Table~\ref{tab:1}, and displayed in the panels (c), (d), (e), and (f) of Fig.~\ref{f:energy}.

\LTcapwidth=\textwidth
\setlength\tabcolsep{10pt}
{\renewcommand{\arraystretch}{1.2}%
\begin{longtable*}{cccccccc}
\caption{\label{tab:1}Initial level energies ($E_i$), $\gamma$ energies ($E_\gamma$), initial and final spins ($I_\text{i}^\pi$ $\to$ $I_\text{f}^\pi$), $\gamma$ intensities ($I^\gamma$), mixing ratios ($\delta$), and adopted multipolarities for the transitions presented in Fig.~\ref{f:level_scheme}. The last two columns present the experimental transition probability ratios for the in-band transitions of DB1 and DB2 as well as for the $\Delta I = 1$ transitions connecting the two bands. The 474.2- and 436.8-keV transitions are tentative and their intensities could not be determined in the present work. All errors reported are purely statistical in nature. For the levels marked with an $\ast$, a negative intensity balance occurs due to missing decay branches from those levels.}\\
\hline\hline
$E_i$ (keV) & $E_\gamma$ (keV) & $I_\text{i}^\pi$ $\to$ $I_\text{f}^\pi$ & $I^\gamma$ (\%) & $\delta$ & Mult. & $\frac{B(M1;\Delta I = 1)}{B(E2;\Delta I = 2)}$ $(\frac{\mu_N^2}{e^2b^2})$ & $\frac{B(E2;\Delta I = 1)}{B(E2;\Delta I = 2)}$ \\\hline
\endfirsthead
\multicolumn{8}{c}%
{\tablename\ \thetable\ -- \textit{Continued from previous page.}} \\
\hline\hline
E$_i$ (keV) & E$_\gamma$ (keV) & $I_\text{i}^\pi$ $\to$ $I_\text{f}^\pi$ & $I^\gamma$ (\%) & $\delta$ & Mult. & $\frac{B(M1;\Delta I = 1)}{B(E2;\Delta I = 2)}$ $(\frac{\mu_N^2}{e^2b^2})$ & $\frac{B(E2;\Delta I = 1)}{B(E2;\Delta I = 2)}$ \\\hline
\endhead
\hline
\endfoot
\endlastfoot
730.7(2) & 372.7(4) & $\frac{15}{2}^- \to \frac{11}{2}^-$ & 100(6) &  & E2 &  &  \\
1390.3(3) & 659.6(4) & $\frac{19}{2}^- \to \frac{15}{2}^-$ & 62(5) & & E2 &  &  \\
1476.8(5) & 746.3(9) & $\frac{17}{2}^- \to \frac{15}{2}^-$ & 8.3(9) & -1.40$_{-0.12}^{+0.14}$ & M1+E2 &  &  \\
2203.1(5) & 811.6(10) & $\frac{21}{2}^- \to \frac{19}{2}^-$ & 3.3(3) & -1.63$_{-0.03}^{+0.02}$ & M1+E2 &  &  \\
 & 726.1(10) & $\frac{21}{2}^- \to \frac{17}{2}^-$ & 2.8(3) & & E2 &  &  \\
2244.4(4) & 854.1(4) & $\frac{23}{2}^- \to \frac{19}{2}^-$ & 30(2) & & E2 &  &  \\
2616.4(4) & 1226.1(4) & $\frac{23}{2}^{(-)} \to \frac{19}{2}^-$ & 3.4(3) & & E2 &  &  \\
2998.6(4) & 754.2(7) & $\frac{25}{2}^- \to \frac{23}{2}^-$ & 1.7(1) & -2.16$_{-0.05}^{+0.06}$ & M1+E2 &  &  \\
 & 795.5(6) & $\frac{25}{2}^- \to \frac{21}{2}^-$ & 2.6(2) &  & E2 &  &  \\
3243.9(4) & 999.5(3) & $\frac{27}{2}^- \to \frac{23}{2}^-$ & 7.4(9) &  & E2 &  &  \\
3487.9(4) & 1243.4(4) & $\frac{27}{2}^{(-)} \to \frac{23}{2}^-$ & 1.6(1) &  & E2 &  &  \\
 & 871.5(5) & $\frac{27}{2}^{(-)} \to \frac{23}{2}^{(-)}$ & 1.9(1) &  & E2 &  &  \\
3862.7(4) & 375.1(7) & $\frac{29}{2}^{(-)} \to \frac{27}{2}^{(-)}$ & 3.0(2) &  & M1+E2 &  &  \\
3954.1(4) & 710.2(5) & $\frac{29}{2}^- \to \frac{27}{2}^-$ & 1.0(1) &  & M1+E2 &  &  \\
 & 955.5(5) & $\frac{29}{2}^- \to \frac{25}{2}^-$ & 2.9(2) &  & E2 &  &  \\
4292.8(4) & 1048.4(7) & $\frac{31}{2}^{(-)} \to \frac{27}{2}^-$ & 2.7(1) &  & E2 &  &  \\
 & 804.9(5) & $\frac{31}{2}^{(-)} \to \frac{27}{2}^{(-)}$ & 0.3(1) &  & E2 &  &  \\
 & 430.1(7) & $\frac{31}{2}^{(-)} \to \frac{29}{2}^{(-)}$ & 1.7(1) & -0.166(4) & M1+E2 &  &  \\
4319.2(4)$^\ast$ & 1075.3(4) & $\frac{31}{2}^- \to \frac{27}{2}^-$ & 2.8(4) &  & E2 &  &  \\
4703.5(4) & 384.8(5) & $\frac{33}{2}^{(-)} \to \frac{31}{2}^-$ &  2.2(1) &  & M1+E2 &  &  \\
 & 840.9(6) & $\frac{33}{2}^{(-)} \to \frac{29}{2}^{(-)}$ & 1.1(1) &  & E2 &  &  \\
 & 410.2(7) & $\frac{33}{2}^{(-)} \to \frac{31}{2}^{(-)}$ & 3.3(2) & $-0.165^{+0.001}_{-0.002}$ & M1+E2 & 12(2) &  2.9(3)\\
5028.3(4) & 735.5(5) & $\frac{35}{2}^{(-)} \to \frac{31}{2}^{(-)}$ & 0.6(1) &  & E2 &  &  \\
 & 324.8(5) & $\frac{35}{2}^{(-)} \to \frac{33}{2}^{(-)}$ & 3.9(2) & -0.159(2) & M1+E2 & 28(5) & 10(2) \\
5065.6(5) & 1111.5(4) & $\frac{33}{2}^{(-)} \to \frac{29}{2}^-$ & 1.1(1) &  & E2 &  &  \\
5110.9(4) & 1156.8(4) & $\frac{33}{2}^{(-)} \to \frac{29}{2}^-$ & 1.0(1) &  & E2 &  &  \\
5162.7(5) & 843.5(5) & $\frac{35}{2}^- \to \frac{31}{2}^-$ & 2.1(5) & & E2 &  &  \\
5374.4(5)$^\ast$ & 308.8(5) & $\frac{35}{2}^{(-)} \to \frac{33}{2}^{(-)}$ & 0.7(1)  &  & M1+E2 &  &  \\
 & 263.3(4) & $\frac{35}{2}^{(-)} \to \frac{33}{2}^{(-)}$ & 0.22(4)&  & M1+E2 &  &  \\
5451.8(4) & 748.8(5) & $\frac{37}{2}^{(-)} \to \frac{33}{2}^{(-)}$ & 0.8(1) &  & E2 &  &  \\
 & 423.6(6) & $\frac{37}{2}^{(-)} \to \frac{35}{2}^{(-)}$ & 3.8(2) & -0.145(3) & M1+E2 & 10(1) & 1.7(2) \\
5709.1(5)$^\ast$ & 598.2(6) & $\frac{37}{2}^{(-)} \to \frac{33}{2}^{(-)}$ & 0.8(1) &  & E2 &  &  \\
 & 335.3(5) & $\frac{37}{2}^{(-)} \to \frac{35}{2}^{(-)}$ & 1.5(1) & -0.049(2) & M1+E2 & 2.6(4) & 0.08(1) \\
5950.5(5)$^\ast$ & 922.2(5) & $\frac{39}{2}^{(-)} \to \frac{35}{2}^{(-)}$ & 0.9(1) &  & E2 &  &  \\
 & 498.2(6) & $\frac{39}{2}^{(-)} \to \frac{37}{2}^{(-)}$ & 2.3(2) & -0.169(2) & M1+E2 & 9(1) & 1.5(2) \\
5996.8(6) & 834.1(4) & $\frac{39}{2}^- \to \frac{35}{2}^-$ & 1.6(4) &  & E2 &  &  \\
6093.9(5) & 384.8(6) & $\frac{39}{2}^{(-)} \to \frac{37}{2}^{(-)}$ & 2.2(1) & -0.150(3) & M1+E2 & 8(2) & 1.8(3) \\ 
& 642.2(6) & $\frac{39}{2}^{(-)} \to \frac{37}{2}^{(-)}$  & 1.0(1) & -0.10(6) & M1+E2 & 0.8(2) & $<$0.07\\
 & 719.6(6) & $\frac{39}{2}^{(-)} \to \frac{35}{2}^{(-)}$ & 0.6(1) &  & E2 &  &  \\
6504.6(5) & 1052.8(4) & $\frac{41}{2}^{(-)} \to \frac{37}{2}^{(-)}$ & 0.7(1) &  & E2 &  &  \\
 & 554.2(6) & $\frac{41}{2}^{(-)} \to \frac{39}{2}^{(-)}$ & 2.4(2) & -0.055(2) & M1+E2 & 18(3) & 0.26(5) \\
6523.2(5) & 814.1(5) & $\frac{41}{2}^{(-)} \to \frac{37}{2}^{(-)}$ & 1.2(1) &  & E2 &  &  \\
 & 428.7(8) & $\frac{41}{2}^{(-)} \to \frac{39}{2}^{(-)}$ & 2.8(2) & -0.127(5) & M1+E2 & 7(1) & 0.9(1) \\
 & 572.9(6) & $\frac{41}{2}^{(-)} \to \frac{39}{2}^{(-)}$ & 0.7(1) & -0.15(3) & M1+E2 & 0.8(1) & 0.07(3) \\
6878.9(6) & 882.1(4) & $\frac{43}{2}^- \to \frac{39}{2}^-$ & 1.4(3) &  & E2 &  &  \\
6982.9(5) & 888.9(6) & $\frac{43}{2}^{(-)} \to \frac{39}{2}^{(-)}$ & 1.3(2) &  & E2 &  &  \\
 & 459.5(6) & $\frac{43}{2}^{(-)} \to \frac{41}{2}^{(-)}$ & 1.2(2) & -0.206(7) & M1+E2 & 3.5(8) & 1.0(2) \\
 & 476.9(6) & $\frac{43}{2}^{(-)} \to \frac{41}{2}^{(-)}$ & 0.5(1) & -0.11(2) & M1+E2 & 1.4(5) & 0.10(4) \\
7027.6(5) & 1077.1(5) & $\frac{43}{2}^{(-)} \to \frac{39}{2}^{(-)}$ & 1.3(1) &  & E2 &  &  \\
 & 522.1(6) & $\frac{43}{2}^{(-)} \to \frac{41}{2}^{(-)}$ & 1.2(2) & -0.138(5) & M1+E2 & 6(1) & 0.6(1) \\
7502.7(5) & 979.5(5) & $\frac{45}{2}^{(-)} \to \frac{41}{2}^{(-)}$ & 0.4(1) &  & E2 &  &  \\
 & 519.5(7) & $\frac{45}{2}^{(-)} \to \frac{43}{2}^{(-)}$ & 2.4(2) & -0.029(5) & M1+E2 & 27(12) & 0.12(6) \\
 & 474.2(6) & $\frac{45}{2}^{(-)} \to \frac{43}{2}^{(-)}$ & -- &  & -- &  &  \\
7594.7(5) & 1090.1(4) & $\frac{45}{2}^{(-)} \to \frac{41}{2}^{(-)}$ & 0.2(1) &  & E2 &  &  \\
 & 567.5(6) & $\frac{45}{2}^{(-)} \to \frac{43}{2}^{(-)}$ & 0.7(1) & -0.059(7) & M1+E2 & 20(12) & 0.3(2) \\
7801.1(7) & 922.2(4) & $\frac{47}{2}^- \to \frac{43}{2}^-$ & 0.5(3) &  & E2 &  &  \\
8033.1(5) & 1050.2(6) & $\frac{47}{2}^{(-)} \to \frac{43}{2}^{(-)}$ & 0.7(1) &  & E2 &  &  \\
 & 530.3(8) & $\frac{47}{2}^{(-)} \to \frac{45}{2}^{(-)}$ & 1.7(2) & -0.076(8) & M1+E2 & 14(4) & 0.4(1) \\
 & 436.8(6) & $\frac{47}{2}^{(-)} \to \frac{45}{2}^{(-)}$ & -- &  & -- &  &  \\
8595.5(5) & 1092.8(5) & $\frac{49}{2}^{(-)} \to \frac{45}{2}^{(-)}$ & 0.3(1) &  & E2 &  &  \\
 & 562.1(6) & $\frac{49}{2}^{(-)} \to \frac{47}{2}^{(-)}$ & 0.8(1) & -0.018(2) & M1+E2 & 16(7) & 0.02(1) \\
8757.9(7) & 956.8(4) & $\frac{51}{2}^- \to \frac{47}{2}^-$ & 1.0(3) &  & E2 &  & \\
\hline\hline
\end{longtable*}}

\begin{figure*}[ht]
  \centering \includegraphics[width=\textwidth]{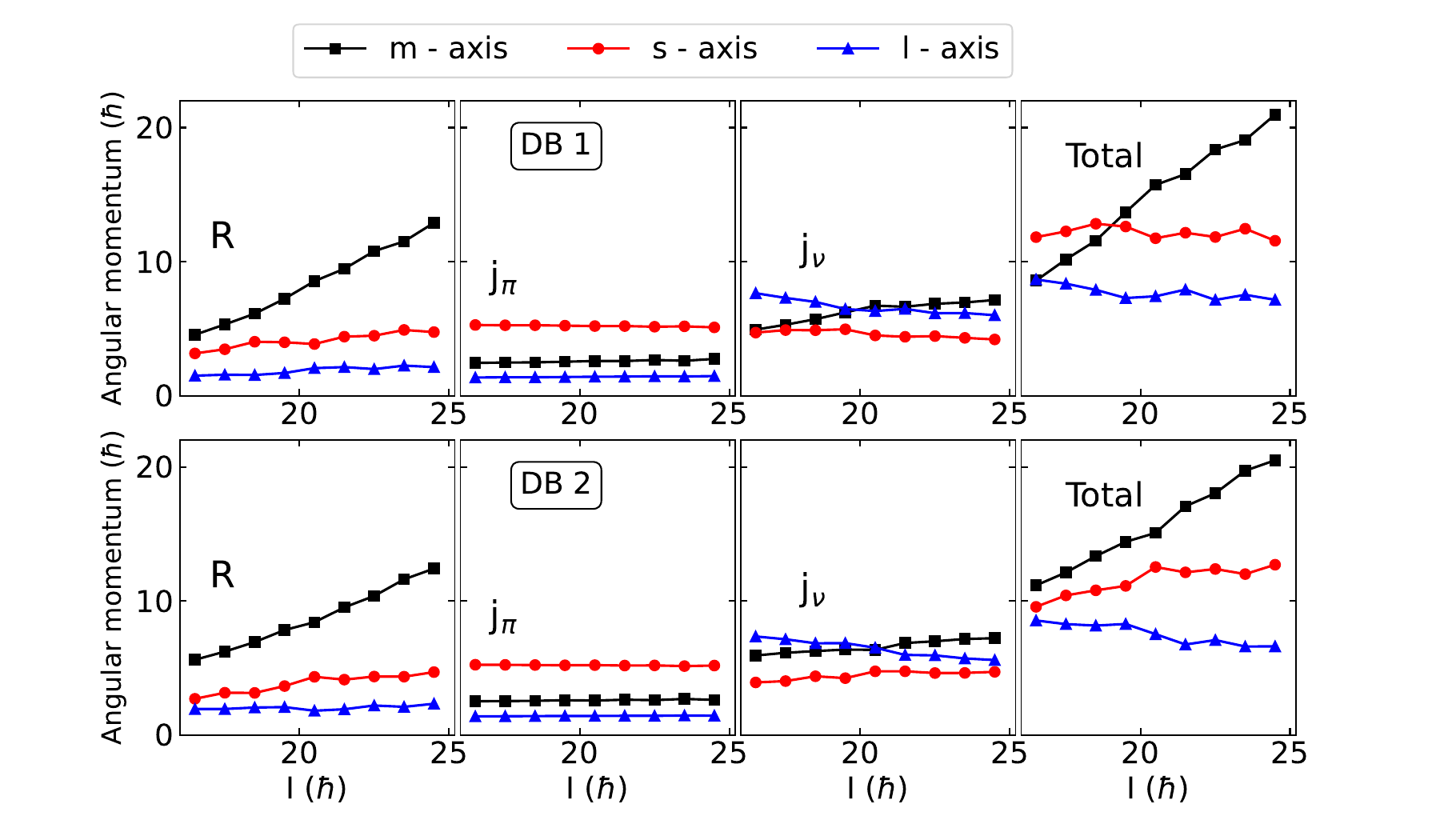}
\caption{(Color online) The root mean square angular momentum components along the medium (m-, black squares),
short (s-, red circles), and long (l-, blue triangles) axes of the rotor $R$, valence proton
$j_\pi$, valence neutron $j_\nu$, and the total angular momentum $J$ for the DB1 band (top panel)
and the DB2 band (bottom panel).}
    \label{f:ang_mom}
\end{figure*}

\begin{figure*}
\centering

\includegraphics[width=\linewidth]{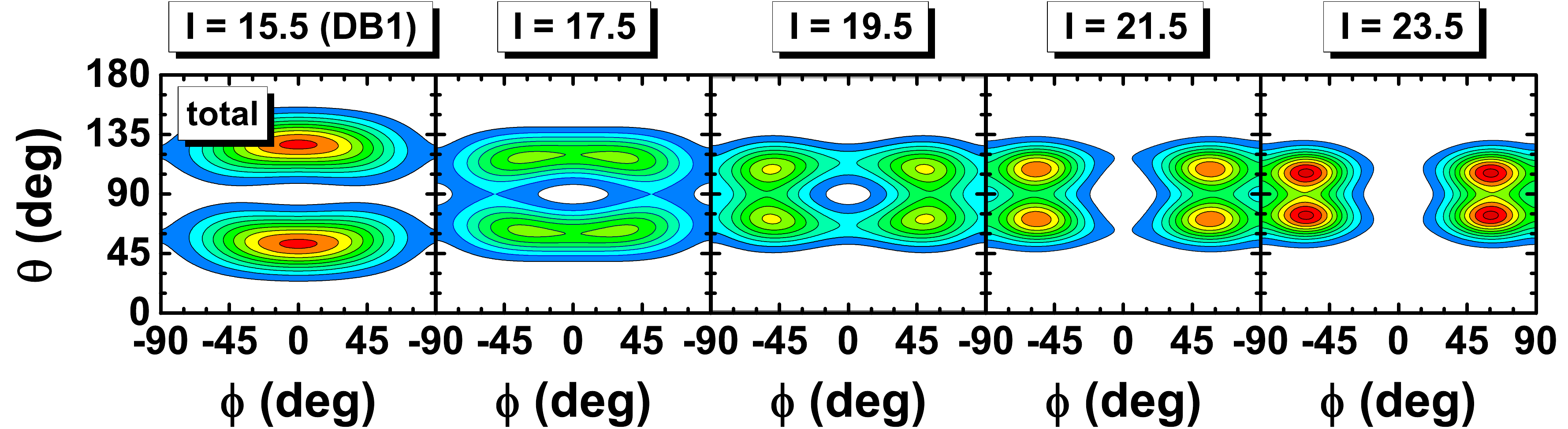}
\includegraphics[width=\linewidth]{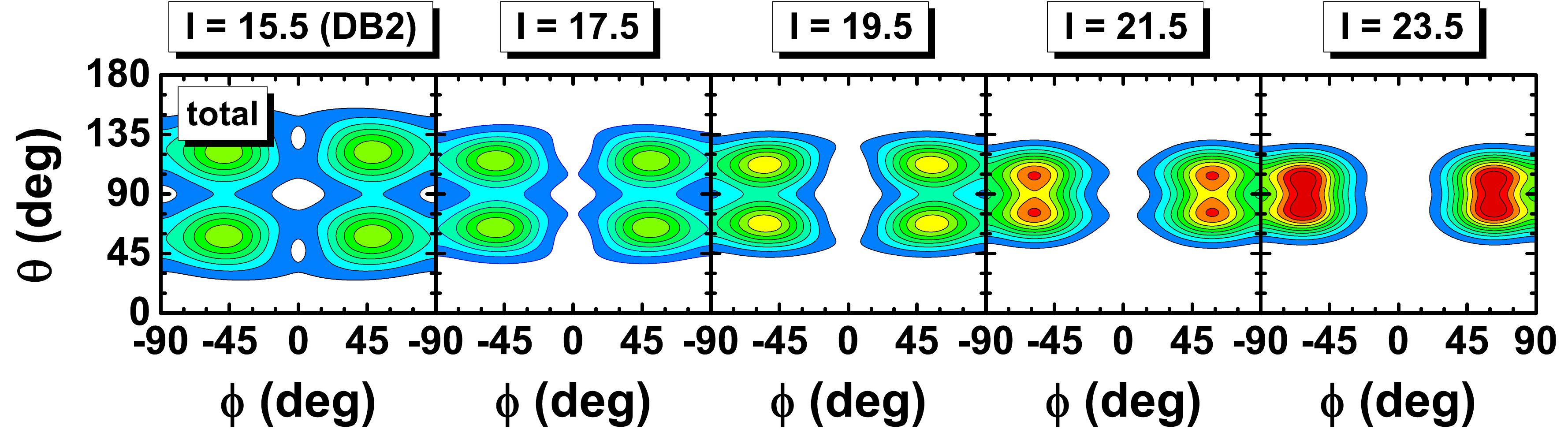}
\includegraphics[width=\linewidth]{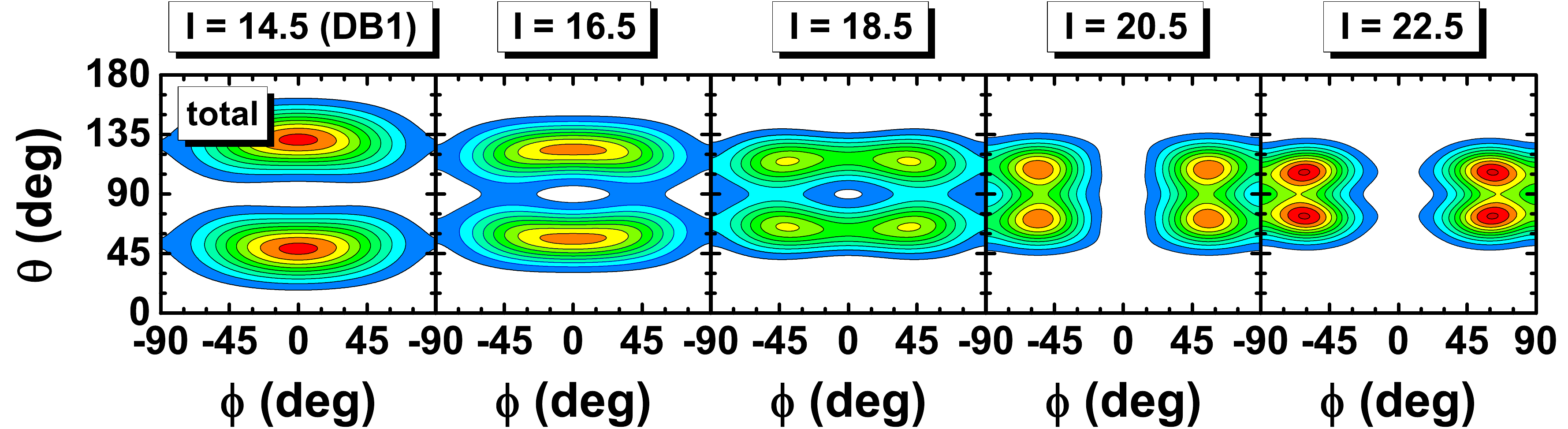}
\includegraphics[width=\linewidth]{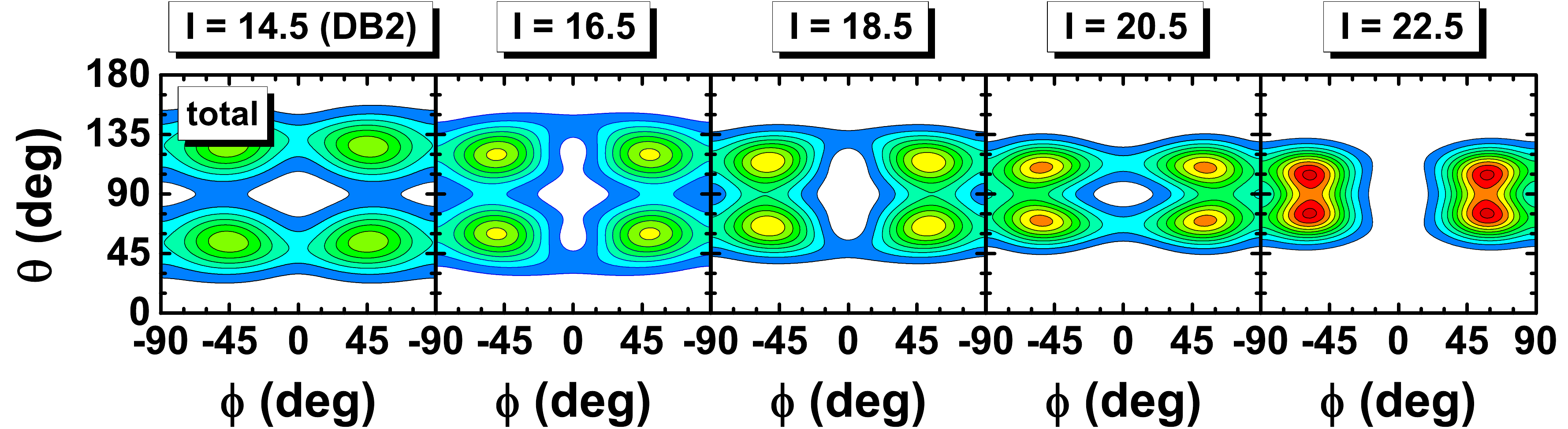}
\caption{\label{f:SCSt}(Color online) Distributions of the probability $\mathcal{P}$($\theta$, $\phi$) for the orientation of the total angular momentum $\bm{J}$  with respect to the body-fixed frame (SCS maps) \cite{Chen22} for the DB1 and the DB2 bands. White, blue, green, yellow, and red are the colors in increasing order of probability.}
\end{figure*}
\begin{figure*}
\centering
\includegraphics[width=\linewidth]{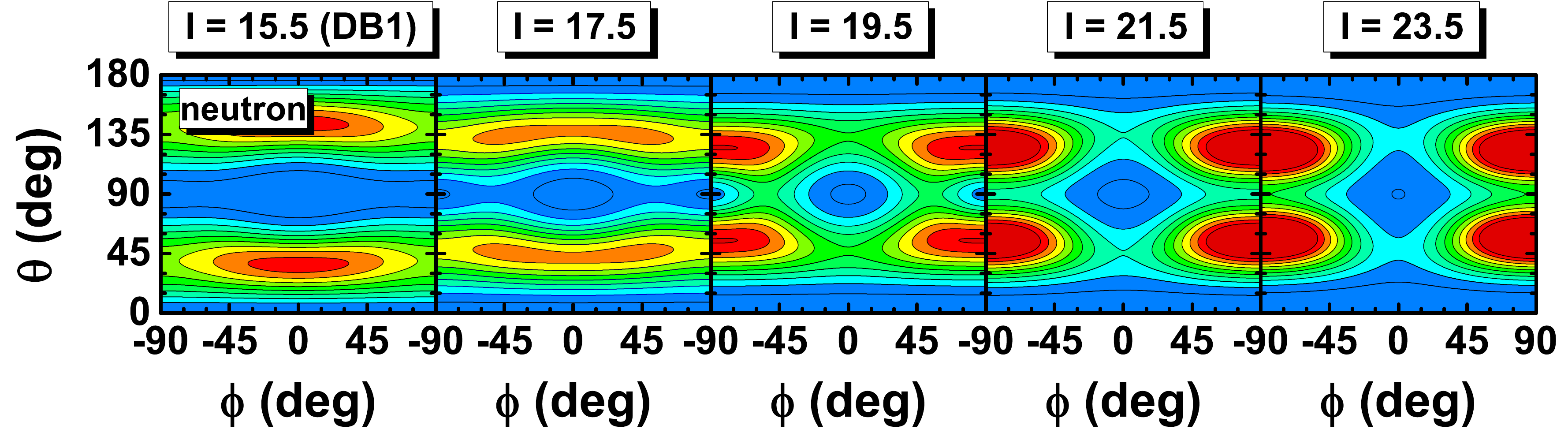}
\includegraphics[width=\linewidth]{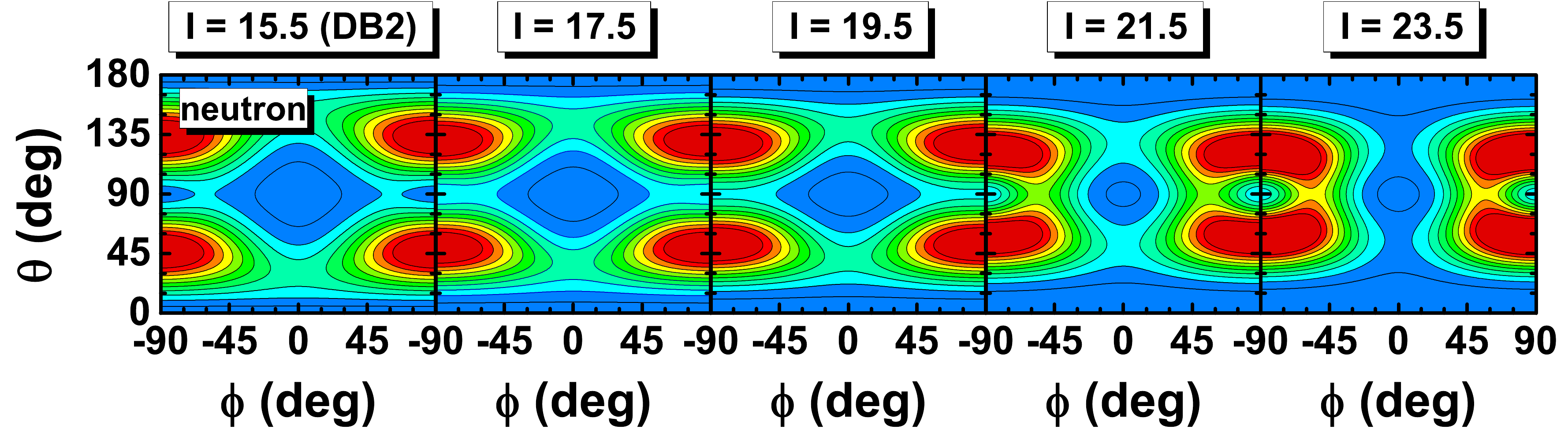}
\caption{\label{f:SCSn}(Color online) Distributions of the probability $\mathcal{P}$($\theta$, $\phi$) for the orientation of the neutron angular momentum $\bm{j}_\nu$ with respect to the body-fixed frame (SCS maps) \cite{Chen22} for the DB1 and the DB2 bands. White, blue, green, yellow, and red are the colors in increasing order of probability.}
\end{figure*}

\begin{figure*}[ht]
\centering
\includegraphics[width=0.8\textwidth]{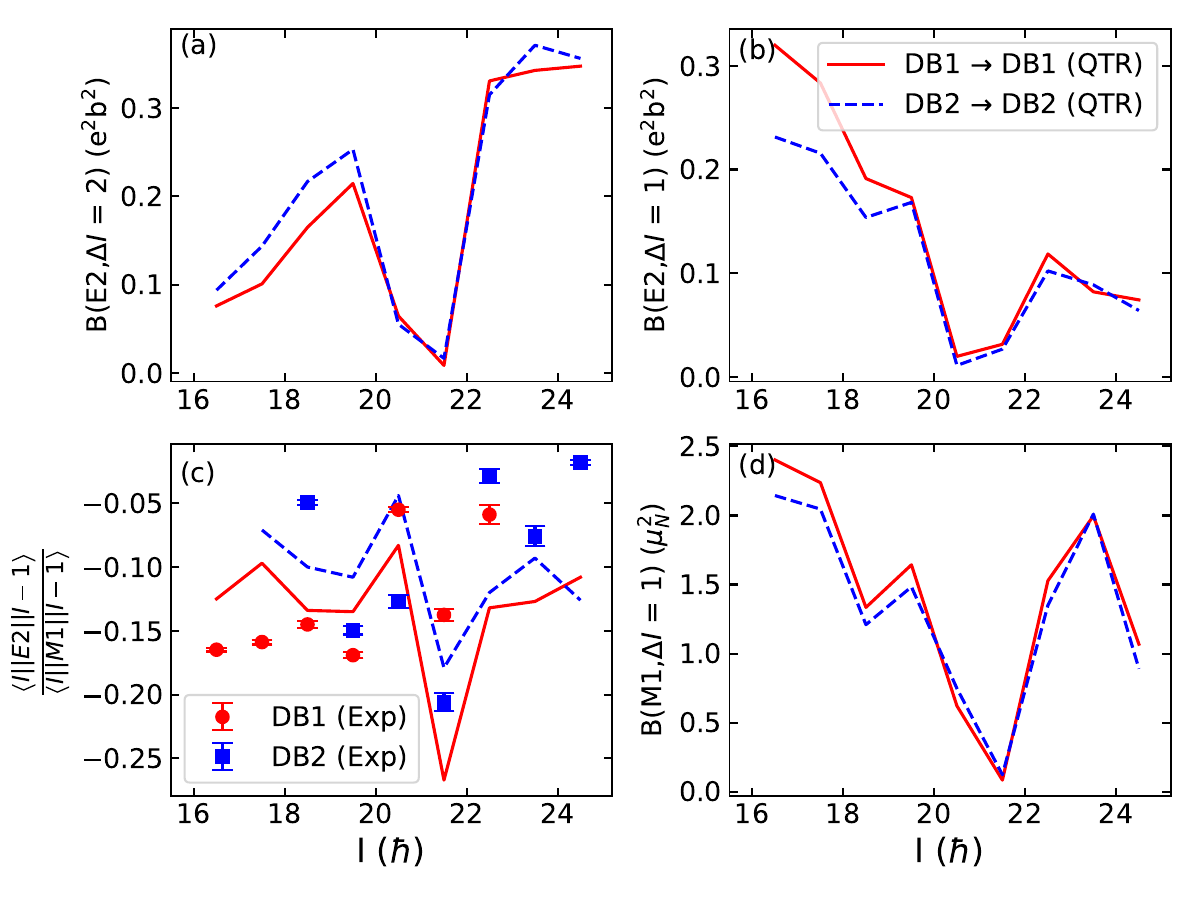}
\caption{\label{f:trans}(Color online) QTR results for the electric quadrupole transition probabilities for the in-band transitions of DB1 and DB2 (a) B(E2; $\Delta I = 2$) and (b) B(E2; $\Delta I = 1$). Panel (c) compares the ratio $\frac{\left<I||E2||I-1\right>}{\left<I||M1||I-1\right>}$ with the experiment. For $I^\pi$ = 47/2$^-$ and 49/2$^-$, the ratio is calculated from the predicted QTR transition energies. The magnetic dipole transition probabilities as predicted by QTR are displayed in panel (d).} 
\end{figure*}

\begin{figure*}[ht]
\centering
\includegraphics[width=0.8\textwidth]{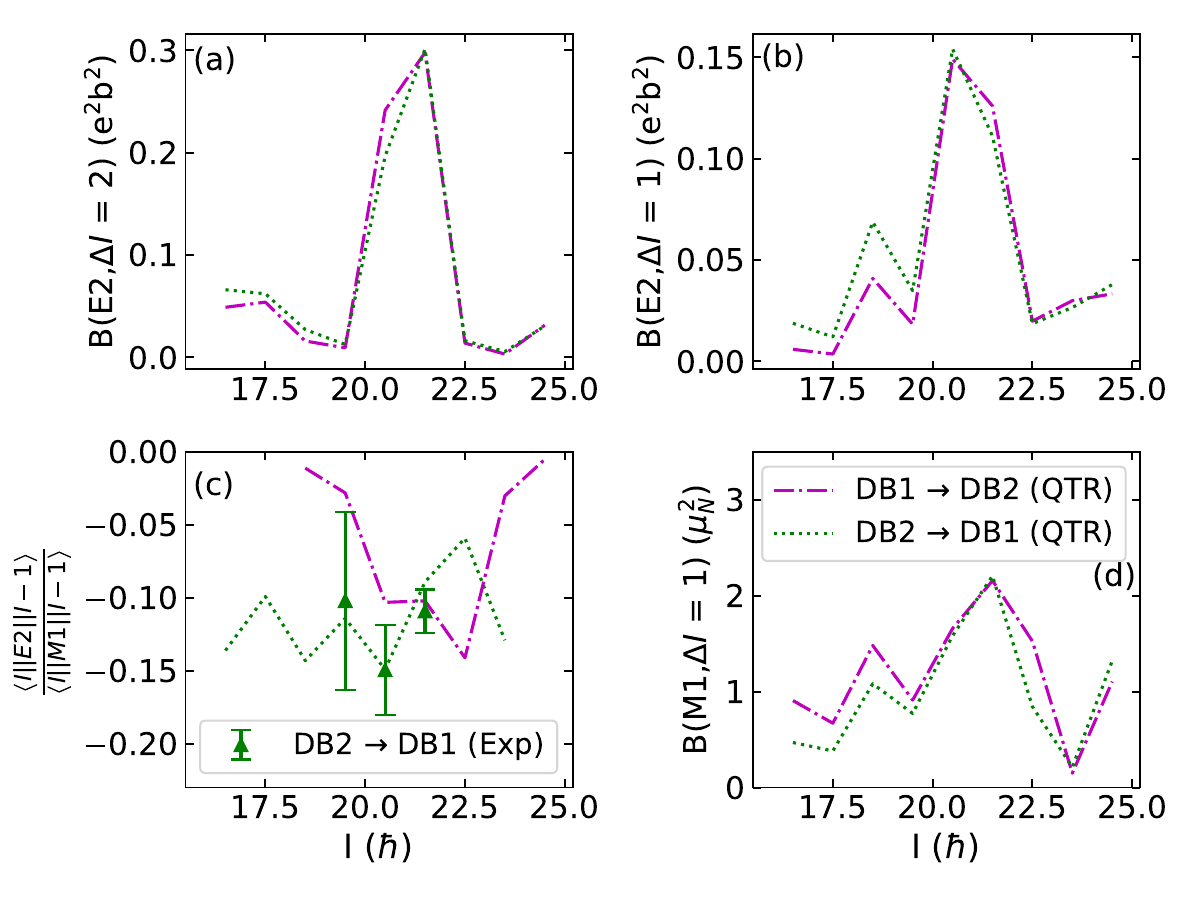}
\caption{\label{f:trans_inter}(Color online) Same as Fig.~\ref{f:trans} but for the $\Delta I$ = 1 connecting transitions between the DB1 and DB2 bands.} 
\end{figure*}

\section{Discussion}
\subsection{Chiral bands in $^{135}$Pr}
 Figure \ref{f:schematic} provides a geometrical representation of the case of the $^{135}$Pr nucleus. The angular momentum of the odd $h_{11/2}$ quasiproton aligned with the short ($s$) axis is combined with a pair of $h_{11/2}$ quasineutrons aligned along the long ($l$) axis. The resulting particle angular momentum vector lies in the $l$-$s$ plane indicated by the black arrow. In the TCV mode the total angular momentum $\bm{J}$ precesses around this axis, i.e., it oscillates between the left- and right-handed octants on the sphere of constant angular momentum. With increasing $J$, the quasiparticle angular momenta $\bm{j}_\pi$ and $\bm{j}_\nu$  follow the motion of $\bm{J}$ to some extend, which increases with $I$ due to the Coriolis force. The TCV evolves into the static CR mode, where $\bm{J}$ spends most of the time in the chiral positions between the $l$-$s$ and the $l$-$m$ planes, while rapidly moving over or tunneling through the potential barriers at the two planes. The stabilization of the chiral 
 geometry at the TCV $\rightarrow$ CR transition is reflected 
 by a change of  $\omega(I)$ from a steeper function to $2A I$.
 Fig.~4 in Ref.~\cite{chiral} illustrates this for the analogue case of $^{134}$Pr. 

The chiral configuration appears through the alignment of the two quasineutrons from the single quasiproton configuration illustrated in Fig.~\ref{f:schematic} (a), which was studied in our previous work~\cite{transverse, 135Pr, two-phonon}. There, we demonstrated the appearance of transverse wobbling (TW), which is yet another manifestation of triaxiality. The present work establishes the existence of chirality in $^{135}$Pr. Although the two identifying signatures of triaxial nuclei have been separately established in the cases mentioned previously, the simultaneous observation of both in the same nucleus, which proves their mutual relation, had not been demonstrated before. 

The presence of chirality has been mainly associated with the appearance of two closely-spaced $\Delta I=1$ bands. Further evidence for CR is the relation,  $I={\cal J}\omega = {\cal J}\left[E(I)-E(I-1)\right]/\hbar$~\cite{chiral} being a straight line emanating from the origin, which is often reformulated as the staggering parameter $S(I)=[E(I)-E(I-1)]/(2I)=1/(2{\cal J})$ being constant~\cite{Koike2003AIP} (see Fig.~\ref{f:energy}(a), for example). Selection rules for the electromagnetic transitions have been derived for the special case of one proton particle and one neutron hole occupying the same high-$j$ orbital and coupled to a triaxial rotor with equal minor moments of inertia~\cite{Koike2004PRL}.

To understand the nature of the motion, we have carried out calculations in the framework of the quasiparticle triaxial rotor model (QTR)~\cite{B.Qi2009PLB, Q.B.Chen2018PLB, Q.B.Chen2020PLB} for the dipole bands DB1 and DB2, based on the unpaired nucleon configuration $\pi(1h_{11/2})^1\otimes \nu(1h_{11/2})^{-2}$. In these QTR calculations, the input deformation parameters $(\beta=0.19, \gamma=22^\circ)$ were determined by the self-consistent adiabatic and configuration-fixed constrained relativistic density functional theory (RDFT)~\cite{J.Meng2006PRC, J.Meng2016book} with the effective interaction PC-PK1~\cite{P.W.Zhao2010PRC}. The pairing correlations in QTR were taken into account by the standard BCS quasiparticle approximation with the empirical pairing gaps $\Delta_\pi=\Delta_\nu=12/\sqrt{A} \approx 1.0~\textrm{MeV}$. Setting  the energy of the spherical $h_{11/2}$ shells equal to zero, the proton chemical potential $\lambda_\pi=-2.09~\textrm{MeV}$ was located at the first level of the $h_{11/2}$ shell, and the neutron chemical potential $\lambda_\nu=0.96~\textrm{MeV}$ was  located 0.3 MeV above the midpoint between the fourth and the fifth levels of the $h_{11/2}$ shell. The ratios between the moments of inertia of the rotor were assumed to be of the irrotational-flow type $\mathcal{J}_k=\mathcal{J}_0\sin^2(\gamma-2k\pi/3)$ with $\mathcal{J}_0=25.0~\hbar^2/\textrm{MeV}$ and $k$ = 1, 2, and 3 corresponding to the $s$, $m$, and $l$ axes. In the calculations of the electromagnetic transitions, the empirical intrinsic quadrupole moment of $Q_0 = (3/\sqrt{5\pi})R_0^2 Z\beta$ with $R_0=1.2A^{1/3}$~fm and the gyromagnetic ratios of $g_R= Z/A=0.44$, $g_\pi(h_{11/2}) = 1.21$, and $g_\nu(h_{11/2}) =-0.21$ were adopted. The results of the QTR calculations are displayed in Figs.~\ref{f:energy} - \ref{f:trans_inter}. 

The right-most panels of Fig.~\ref{f:ang_mom} display the expectation 
values $J_i=\sqrt{\langle J_i^2\rangle}$ of the total
angular momentum components with respect to the body-fixed axes. For $I\geq 41/2$, 
the spin where DB1 and DB2 come closest, $J_s\approx 12\hbar$ and $J_l\approx 8\hbar$, while
$J_m$ continuously increases. 
This is the angular momentum geometry that characterizes chiral rotation  as sketched in 
Fig.~\ref{f:schematic} (c). The other panels illustrate 
how the geometry comes about. The $l$-component is generated by the $h_{11/2}$ quasineutron
 pair. The $s$-component is the sum of the $h_{11/2}$ quasiproton and quasineutron contributions.
 The $m$-component is mainly generated by the collective angular momentum $R$, with an 
 increasing contribution from the $h_{11/2}$ quasineutron pair. The real angular momentum 
 composition deviates from the sketch in Fig.~\ref{f:schematic}, because the quasineutron pair
 does not have the pure hole character of the schematic figure and reacts to the 
 Coriolis force by reorienting toward the $m$-axis. 
 
Figures~\ref{f:SCSt} and \ref{f:SCSn} complement the mean values of the 
angular momentum components shown in Fig.~\ref{f:ang_mom}. Respectively, they 
display the probability distributions of the total angular momentum $\bm{J}$ vector  and of the quasineutron angular momentum vector  $\bm{j}_\nu$ as functions of their angles with respect to the body-fixed frame, where $\theta$ is the angle with the $l$-axis and $\phi$ is the angle of the projection on the $s$-$m$ plane with the $s$-axis. The distributions represent the classical orbits, as sketched in Fig.~\ref{f:schematic}, where their fuzziness represents the quantum mechanical limit of the concept of classical orbits. The strict definition of these Spin Coherent State (SCS) maps can be found in Ref.~\cite{Chen22}.

Figure~\ref{f:SCSt} presents the chiral angular momentum arrangements, which are  schematically illustrated in  Fig.~\ref{f:schematic}. 
For $I$ = $29/2$, $31/2$ and $33/2$, DB1 has the character of the $n$ = 0 state of an anharmonic TCV with a distribution that corresponds to a squeezed version of the ellipse in Fig.~\ref{f:schematic} (b) and its reflection through the $\theta=90^\circ$ plane. The distributions of the DB2 partner band are centered at the left- and right-handed angular momentum arrangements as characteristic for the CR regime.


The SCS plots for $I = 11.5$, $12.5$, and $13.5$ in DB1 are not shown. As expected for $n=0$ TCV states, they resemble the distribution observed for $I = 14.5$ in DB1, with maxima located at the same positions but appearing narrower, indicating smaller zero-point fluctuations. In contrast to the analogous case of $^{134}$Pr (see Fig.~3 of Ref.~\cite{CF2025}), the DB2 states for these spins—also not shown—do not exhibit the anticipated $n=1$ TCV pattern characterized by a broader elliptical shape, but instead correspond to configurations with a different relative orientation of the two quasineutrons. These excited states were not analyzed further, as they have not been experimentally observed.

 For $I\geq 35/2$  the distributions of DB1  become centered at the left- and right-handed angular momentum arrangements. The distributions of DB2
continue to be  arranged in this way for $I\geq 35/2$. The four maxima of DB1 and DB2 gradually move towards the $s$-$m$ plane. For $I\geq 41/2$ the distributions near the 
four maxima (on the front hemisphere) of DB1 and DB2 are almost identical, that is,   the CR structure is retained. The bands differ in the way the maxima are coupled to each other. The regions $\phi<0^\circ$ and $\phi>0^\circ$ of DB1 and DB2 separate from each for $I=41/2$ and $I=35/2$, respectively. 

The probability distribution of the angular momentum of the one-quasiproton configuration (not shown) stays well centered at the $s$-axis. The angular momentum distribution of the two-quasineutron configuration in Fig.~\ref{f:SCSn} has a substantial  $l$-component, as outlined in Figs.~\ref{f:schematic} (b) and (c). With  increasing $I$, the vector  $\bm{j}_\nu$ tilts toward the $s$-$m$ plane as indicated by the decreasing $l$-component and increasing $s$- and $m$-components. 

 Figure~\ref{f:energy}(a) compares the QTR energies with the experiment. 
 Although there is a good overall agreement, the QTR energies 
 exhibit a signature staggering for $I\geq 39/2$.
 For $\alpha=-1/2$ DB1 and DB2 repel each other while for 
 $\alpha=1/2$ they are close together, which is not seen in the experiment. The staggering correlates with the coupling of
 the maxima in Fig.~\ref{f:SCSt}. The height of the saddles at $\theta=90^\circ$ differ more for $\alpha=-1/2$ than for $\alpha=1/2$.
 
 Remarkably, the QTR model reproduces the very small energy difference of less than 20 keV between the $I$ = $41/2$ levels. The distributions of DB1 and DB2 in Fig.~\ref{f:SCSt} become very similar around $I$ = $41/2$ and remain so for larger $I$. The small difference indicates that the two states must have opposite intrinsic symmetry with respect to the chirality-changing transformation. The chiral-partner bands based on the $\pi h_{11/2} \otimes \nu h_{11/2}$ configuration in the neighboring isotope $^{134}$Pr \cite{Starosta2001PRL,Tonev2006PRL,Tonev2007PRC,timar2011} show the same crossing with an energy difference of about 30 keV between the $I$ = $15$ levels. 
 
 The analog situation for octupole vibrations has been discussed in Ref.~\cite{Frauendorf08}. The one-phonon band has a rotationally aligned angular momentum of about 3$\hbar$ and odd $I$, which leads to the crossing of the even-$I$ zero-phonon band. Mixing cannot appear because the bands have different spin and parity. 
 
 As seen in Fig.~\ref{f:energy}(b), DB2 has a surplus angular momentum of about 2$\hbar$ with respect to DB1, which generates the band crossing. The very small difference in the energy between the $I$ = $41/2$ levels indicates an opposite symmetry with respect to an intrinsic transformation for the two states with the same parity and spin. This can only be the change between the left- and right-handed arrangements of the three angular momentum constituents. The details of the band-crossing phenomenon and their relation to the avoided crossing seen in the TW configuration of $^{135}$Pr (see Ref.~\cite{Chen22}) are yet to be investigated. 
 
 The experimental values of $\omega(I)$ presented 
 in Fig.~\ref{f:energy}(b) give an average  difference of about 2$\hbar$ of aligned angular momentum between DB2 and
 DB1, which leads to the characteristic crossing transverse chiral vibrational bands~\cite{Q.B.Chen2018PRC_v1, Q.B.Chen2019PRC}. The QTR values show irregularities which are not seen in the experiment. The average difference  of aligned angular momentum between DB2 and
 DB1 amounts to about 1$\hbar$, which is reflected by 
 their smaller distance in Fig.~\ref{f:energy}(a).
 
 As mentioned before, the approach of the experimental $\omega(I)$ values to the relation $\omega= 2A I$ at large angular momenta has been interpreted as a signal for the transition from the TCV to the CR regime. As illustrated in
 Fig.~\ref{f:schematic} (c), in the CR regime 
 $E(I)=A\left(I(I+1)-I_\perp^2\right)$ with $I_\perp$ being 
 the angular momentum in the $l$-$s$ plane, 
 for which $\omega=2A(I+1/2)$. 
 The average QTR values do approach  this limit due to
 the continuous increase of the quasineutron $m$-component seen in Figs.~\ref{f:ang_mom} and ~\ref{f:SCSn}. Nevertheless, the SCS plots 
 in Fig.~\ref{f:SCSt} indicate an early appearance of the CR regime. 
 The discrepancy suggests that the QTR overestimates the reorientation
 of the quasineutron pair, which would also explain the signature
 staggering of the bands that is not seen in the experiment.

The QTR values for the $B(E2)$ and $B(M1)$ probabilities for the DB1 and DB2 bands displayed in Fig.~\ref{f:trans} are very similar, which is expected for chiral-partner bands. The respective values for interband transitions are presented in Fig.~\ref{f:trans_inter}. With the adopted association of the states within DB1 and DB2, the intraband $B(E2,I\rightarrow I-2)$ values for $I$ = $41/2$ and $43/2$ are strongly reduced, while the interband values are strongly enhanced, such that their sum smoothly increases with $I$. The $B(E2,I\rightarrow I-1)$ values behave in the same way, while their sum smoothly decreases with $I$. The intraband and interband $B(M1,I\rightarrow I-1)$ values (see Figs.~\ref{f:trans}(d) and \ref{f:trans_inter}(d)) have, respectively, a minimum and maximum at $I$ = $43/2$, while their sum slightly decreases with $I$.

The QTR mixing ratios for the intraband transitions in Fig.~\ref{f:trans}(c) follow the experimental ones fairly well. The experimental differences between DB1 and DB2 are larger and opposite compared to the QTR calculations. The QTR values for the interband transitions of $\left<E2\right>/\left<M1\right> \simeq -0.12$ (Fig.~\ref{f:trans_inter}(c)) are close to the experimental ones as well. 

From Fig.~\ref{f:energy}(e), it is observed that the experimental and theoretical intraband $B(M1,I\rightarrow I-1)/B(E2, I\rightarrow I-2)$ ratios for the DB1 and DB2 bands are similar, supporting the chiral-partner band interpretation. With the above-mentioned association of DB1 and DB2 with the lowest QTR states, the calculated ratios reproduce the experimental ones fairly well, which also holds for the intraband $B(E2,I\rightarrow I-1)/B(E2,I\rightarrow I-2)$ values, displayed in Fig.~\ref{f:energy}(f). The fact that the $B(E2,I\rightarrow I-1)$ values are comparable with the $B(E2,I\rightarrow I-2)$ ones is another indication of chiral geometry. The experimental ratios of DB1 and DB2 differ more from each other than predicted by the QTR. The discrepancies reflect the differences of the $\left<E2\right>/\left<M1\right>$ ratios in Figs.~\ref{f:trans}(c) and \ref{f:trans_inter}(c). Further, they may be related to differences of the $B(E2,I\rightarrow I-2)$ values between the chiral-partner bands such as those seen in the neighboring nucleus $^{134}$Pr, which have been attributed to a coupling of the orientation degrees of freedom to the shape ones~\cite{Tonev2007PRC}. 

Figure~\ref{f:energy}(d) compares the experimental and theoretical values of the $B(E2, I\rightarrow I-1)_{\textrm{out}}/B(E2, I\rightarrow I-2)_{\textrm{in}}$ probability ratios for the $\Delta I$ = 1 transitions connecting DB2 with DB1. Fig.~\ref{f:energy}(c) presents the same comparison for the $B(M1, I\rightarrow I-1)_{\textrm{out}}/B(E2,I\rightarrow I-2)_{\textrm{in}}$ ratios. 
The QTR energy difference between DB2 and DB1 is less than 150 keV for $I$ = $39/2$, $41/2$, and $43/2$. A possible non-diagonal matrix element of the order of 100 keV missed by the QTR model Hamiltonian will cause a superposition of the QTR states in DB1 and DB2 that will redistribute the $B(E2)_{\textrm{in}}$ and $B(E2)_{\textrm{out}}$ values such that their sum remains the same as seen in Figs.~\ref{f:trans}(a) and (b). The same holds for the $B(M1)_{\textrm{in}}$ and $B(M1)_{\textrm{out}}$ values as seen in Figs.~\ref{f:trans}(d) and \ref{f:trans_inter}(d), respectively. Such a superposition could account for the differences observed in Figs.~\ref{f:energy}(c) and (d). However, it would not change the QTR $\left<E2\right>/\left<M1\right>$ ratios much, which are similar for the interband and intraband transitions (see Figs.~\ref{f:trans}(c) and \ref{f:trans_inter}(c)). Clearly, lifetime measurements would be essential to establish the absolute reduced transition probabilities, and to remove the present ambiguities of being restricted to only ratios.  

As observed in Fig.~\ref{f:trans_inter}, the QTR calculations predict the same reduced transition probabilities for the DB1 $\rightarrow$ DB2 transitions as for the DB2 $\rightarrow$ DB1 ones, which is expected for chiral partners. The absence of DB1 $\rightarrow$ DB2 transitions in the experiment seems to suggest that the two bands exhibit chirality that is more akin to a zero- and one-quantum vibrational state than the static chirality predicted by the QTR model. This is in accordance
with the above discussed deviations of the QTR function $\omega(I)$ from the 
experimental one.

The neighboring nucleus $^{134}$Pr has been studied in detail in 
Refs.~\cite{Starosta2001PRL, Tonev2006PRL, Tonev2007PRC, timar2011}. Compared to $^{135}$Pr, it has only one
hole-like quasineutron combined with the particle-like quasiproton.  Its angular momentum
geometry is closer to the ``ideal" arrangement of a proton particle, a neutron hole  
and a $\gamma=30^\circ$ rotor core, for which the authors of Ref.~\cite{Koike2004PRL}
derived selection rules, which generate the characteristic staggering of the 
intra and inter band $M1$ transitions. In Ref.~\cite{Zhang2007}, the authors demonstrated that
the staggering is quickly quenched when the deformation deviates from $\gamma=30^\circ$.
 In Ref.~\cite{Tonev2007PRC}, the authors interpreted their data for $^{134}$Pr in terms of a model that couples the proton particle and the neutron hole
  with an Interacting Boson Model (IBM) core and compared it with the QTR model. The fluctuations  of the triaxial 
  shape, which the IBM core takes into account, removed the staggering pattern and
  accounted for the difference between  the intra band $B(E2)$ values of the two partner bands.
  Nevertheless, the analysis of the angular momentum geometry pointed to the presence of
  ``weak dynamical chirality".

 With one proton particle and two neutron holes, $^{135}$Pr is asymmetric, which violates 
 prerequisites for the selection rules derived in Ref.~\cite{Koike2004PRL}. 
 As seen in Figs.~\ref{f:trans} and \ref{f:trans_inter}, the decay pattern changes to 
 strong intra band  and weak inter band $E2$ and $M1$ transitions away from the region 
 where the partner bands cross. In the crossing region, strong inter band 
 transitions arise, where the sum of the inter band and intra band values for $B(E2)$ as well as
 $B(M1)$ changes smoothly with $I$. The same pattern has been found in the analog asymmetric case
 of $^{135}$Nd, with two proton particles and one neutron hole~\cite{zhu,Mukhopadhyay2007PRL}.
 The data were well reproduced by the QTR calculations in Ref.~\cite{B.Qi2009PLB},
 where the analysis of the angular momentum geometry demonstrated the 
 chiral nature of the two bands. Lifetime measurements~\cite{Mukhopadhyay2007PRL}
 demonstrated that in contrast to $^{134}$Pr, the two bands in $^{135}$Nd have very similar
 stretched intra band $E2$ transitions, indicating a more stable shape. One may
 speculate that this is the case for the three quasiparticle configuration in $^{135}$Pr as well.

\begin{figure}
    \centering
    \includegraphics[width=\columnwidth]{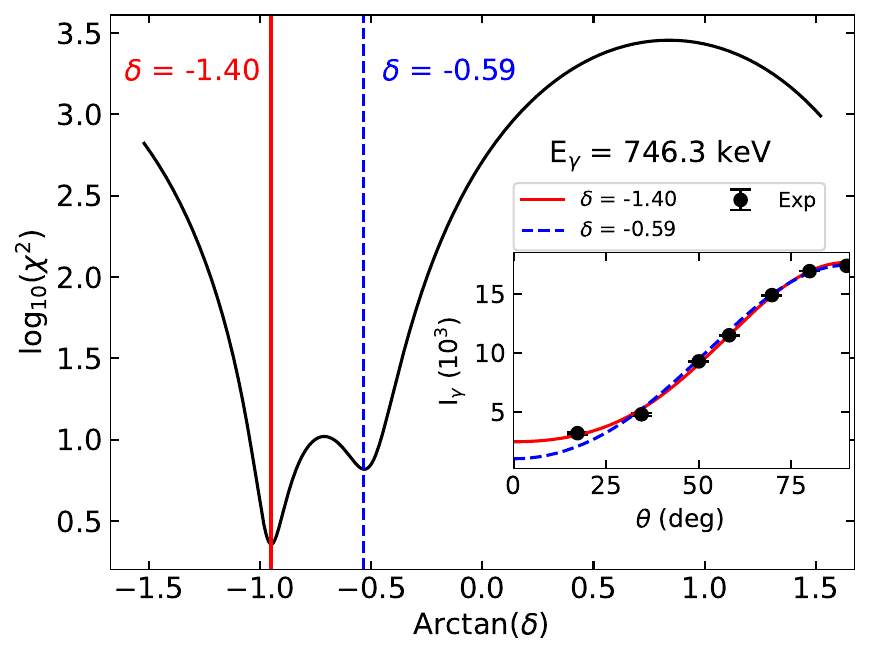}
    \includegraphics[width=\columnwidth]{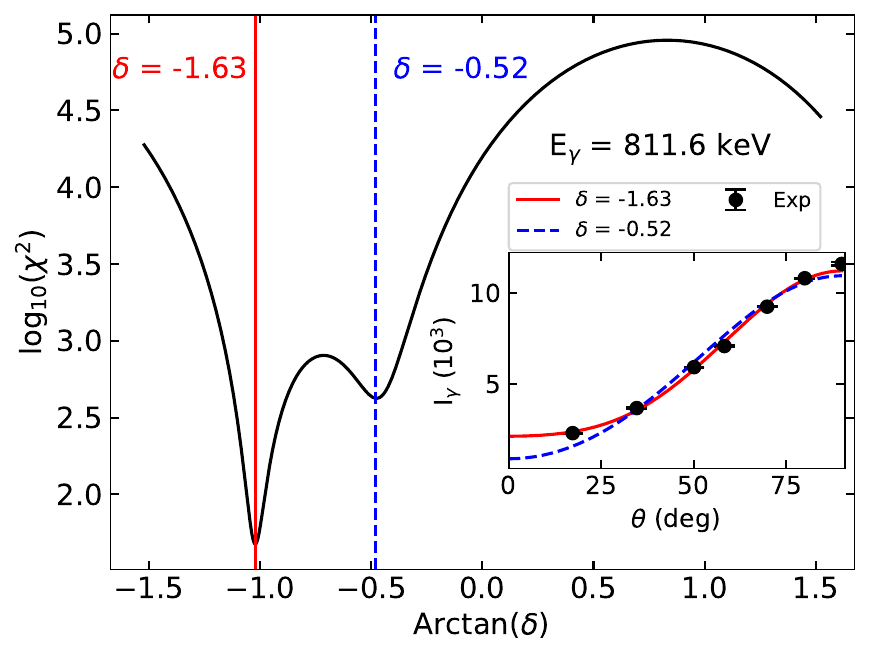}
    \includegraphics[width=\columnwidth]{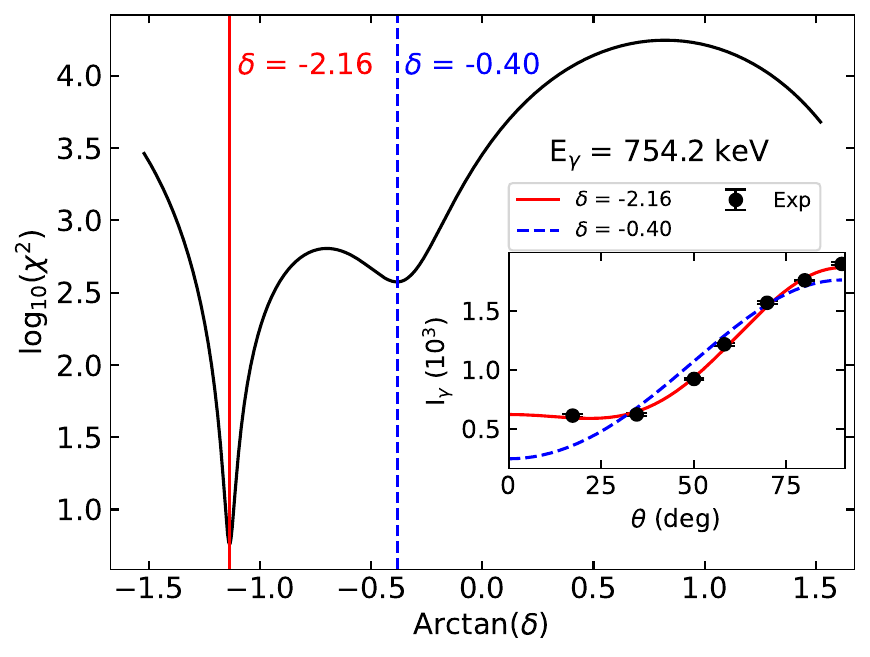}
    \caption{(Color online) Calculated $\chi^2$ values between the theoretical and experimental angular distributions of the 746-, 812-, and 754-keV transitions in $^{135}$Pr (Ref.~\cite{135Pr,two-phonon}) as a function of the mixing ratio $\delta$; the ratio $\sigma/I$ adopted is 0.2. (Insets) Experimental angular distributions of the 746-, 812-, and 754-keV transitions (black full circles), and calculated distributions with the adopted $\delta$ (red solid line) and the other $\delta$ corresponding to a higher $\chi^2$ (blue dashed line).}
    \label{fig:chi-square}
\end{figure}

\subsection{Refutation of Lv \textit{et al.}}

In a recent paper, Lv \textit{et al.}~\cite{lv} questioned our earlier work~\cite{135Pr,two-phonon}, which reported on the observation of TW in $^{135}$Pr. As an important result of the present study -- the simultaneous observation of transverse wobbling and chirality -- relies on the angular distribution results, in the following we refute their criticisms -- both experimental and theoretical --  and present additional evidence to further strengthen our arguments. 

Ref.~\cite{lv} claims that the mixing ratios, $|\delta |$, for the $\Delta I = 1$ transitions connecting the wobbling bands could also be $<$ 1, thereby questioning the $E2$ nature of these transitions and, hence, the wobbling interpretation of the level sequences. Their argument is that the angular distributions provide two $\delta$ values, both of which are equally likely. This observation is in direct contrast to the mixing ratios reported in Refs.~\cite{135Pr,two-phonon}. We assert that their claim is erroneous and unfounded. In our measurements, performed with the Gammasphere array, angular distribution information could be obtained at sufficient angles so as to reliably extract both the $A_2$ and $A_4$ coefficients of the standard angular distribution expression. This allowed for the mixing ratios to be determined unambiguously by differentiating between the two possible local $\chi^2$ minima, with one of them consistently providing the better fit as determined by the standard $\chi^2$-minimization techniques. The calculated $\chi^2$ values between theoretical and experimental angular distribution for the 746-, 812-, and 754-keV transitions connecting the wobbling bands as a function of the mixing ratios $\delta$ are presented in Fig.~\ref{fig:chi-square}. Clearly, the data gives only one true minimum in each case, corresponding to the $|\delta|>1$ value (also reported in Ref.~\cite{135Pr}) for these transitions. In the insets, we have presented the experimental angular distributions for these transitions along with the calculated distributions corresponding to the two $\delta$ values obtained from solving the quadratic equation (See Eq. 2.46 in Ref.~\cite{springer_book}). Just looking at this figure could lead to an erroneous conclusion, as Lv et al. \cite{lv} appear to have reached, that the two fits are ``similar''; however, the $\chi^2$ minimization process unambiguously gives only one value. The results provided in Refs.~\cite{135Pr,two-phonon}, thus, clearly establish the predominantly $E2$ character of the transitions connecting the wobbling bands, rendering firm evidence for the occurrence of wobbling motion in this nucleus. 

Similar critical comments have been made in Ref.~\cite{lv} about the polarization asymmetries reported in Ref. \cite{135Pr}. We wish to point out that the polarization asymmetry measurements were provided merely as a further confirmation of the high $E2$ admixtures established by the angular distribution data, and \textit{not} as a way to determine the mixing ratios therefrom. For the two transitions where it was possible to extract asymmetry parameters, those are positive even beyond the limits of the uncertainties. As reported in our paper, the asymmetry parameters for the transitions identified as ``predominantly $E2$'' are all positive; are closer to those for the known pure $E2$ transitions from the yrast band; and have a sign opposite to that obtained for the 594-keV transition from the signature-partner band to the yrast sequence, which has been designated as ``primarily $M1$''. Indeed, our procedure, and the associated arguments, are identical to those used in the first reports of wobbling motion in nuclei \cite{odegard,Jensen,amro}: in the  case of the Lu isotopes as well, large negative mixing ratios were found and limited polarization results were employed to confirm the ``electric'' nature of the wobbling transitions.
\begin{figure}[t]
\centering
\includegraphics[width=\columnwidth]{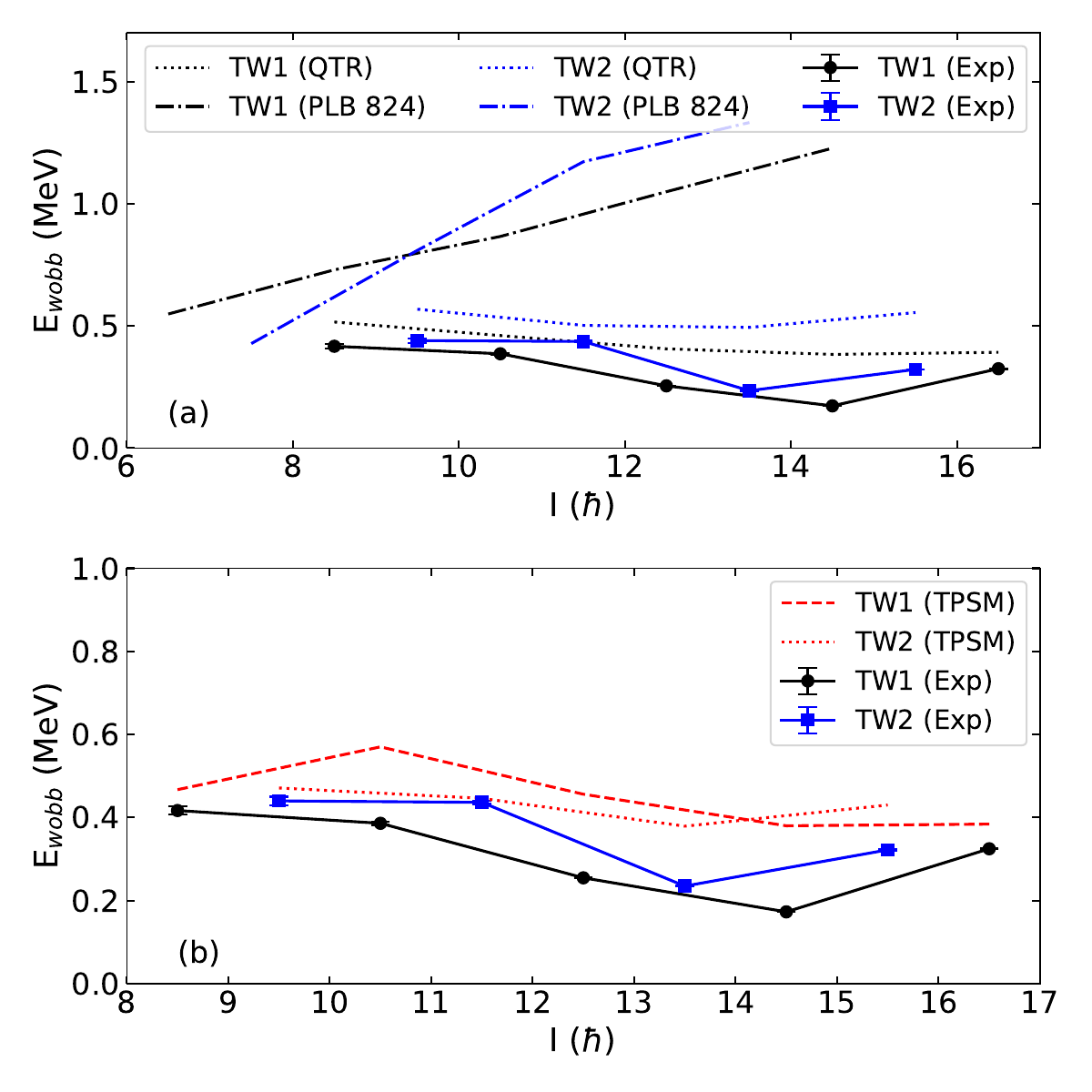}
\caption{\label{f:Lvenergy}(Color online) Wobbling energies, E$_{wobb}$, as a function of spin for the TW1 and TW2 bands in $^{135}$Pr. The experimental values are given as black circles (TW1) and blue squares (TW2). Also shown for comparison are the QTR values from Ref.~\cite{two-phonon,lawrie} (black and blue dotted lines) and that from Ref.~\cite{lv} (black and blue dashed-dotted lines) in the upper panel, and the corresponding TPSM values (dotted and dashed-dotted red lines) in the lower panel.} 
\end{figure}

The authors of Ref.~\cite{lv} also wrongly claim that, in the theoretical analyses of Refs.~\cite{transverse, 135Pr, two-phonon}, the frozen alignment approximation was applied. We want to underline that the experimental results were compared with the QTR calculations, which fully account for the re-orientation of the particle angular momentum. The main difference between the QTR calculations of Ref.~\cite{lv} and Refs.~\cite{transverse, 135Pr, two-phonon} consists in the ratios between the moments of inertia. The calculations in Ref.~\cite{lv} assume irrotational-flow ratios. The large ratio of four between the $m$- and $s$-axes generates an early realignment of the total angular momentum with the $m$-axis. As a consequence, the energy differences between the lowest band and the next two bands increase with angular momentum, in contrast to the experimental energies, as seen in the upper panel of Fig.~\ref{f:Lvenergy}. 

In the calculations of Refs.~\cite{135Pr,two-phonon}, a ratio of about two is assumed, which leads to a later reorientation of the total angular momentum. As a consequence, the relative energies first decrease and then increase, which is the signature of transverse wobbling. As seen in the upper panel of Fig.~\ref{f:Lvenergy}, the angular momentum dependence of the relative energies agrees qualitatively with the experimental ones. The ratio of 2 between the moments of inertia was adjusted such that the QTR energies come as close as possible to the experimental ones. 

The Triaxial Projected Shell Model (TPSM) calculations in the lower panel of Fig.~\ref{f:Lvenergy} evaluated the rotational response microscopically. The similarity between the QTR and TPSM results lends additional credibility to the adjusted ratio of two.

\section{Conclusion}
In summary, new experimental evidence supporting the interpretation of a previously observed pair of bands in $^{135}$Pr as chiral partners based on the $\pi(1h_{11/2})^1\otimes \nu(1h_{11/2})^{-2}$ configuration has been established using a high-statistics Gammasphere experiment with the $^{123}$Sb($^{16}$O,4n)$^{135}$Pr reaction. The predominant magnetic dipole character of the $\Delta I=1$ transitions between the two bands has been demonstrated by the precise measurement of angular distributions. Results from QTR calculations are in fair agreement with the experimental observations and support the chiral-partner interpretation. With this observation, both signatures of triaxiality in nuclei—chirality and wobbling—have been evidenced in the same nucleus, placing the triaxial nature of nuclear shapes on a firmer than ever footing.

\begin{acknowledgments}

This work has been supported in part by the U.S. National Science Foundation [Grants No.~PHY-1713857 (UND), No.~PHY-2011890 (UND), No.~PHY-2310059 (UND)  and No.~PHY-190740 (USNA)]; by the U. S. Department of Energy, Office of Science, Office of Nuclear Physics [Contract No.~DE-AC02-06CH11357 (ANL), Grants No.~DE-FG02-95ER40934 (UND), No.~DE-FG02-97ER41033 (UNC), and No.~DE-FG02-97ER41041 (TUNL)]; and by the National Natural Science Foundation of China under Grant No.~12205103. This research used resources of ANL's ATLAS facility, which is a DOE Office of Science User Facility.

\end{acknowledgments}

\bibliography{chiral_wobb}

@book{bohr,
    author    = "A. Bohr and B. R. Mottelson",
    title     = "Nuclear Structure",
    year      = "1975",
    publisher = "W. A. Benjamin",
    address   = "New York",
    Volume    = {II},
    url	      = "https://books.google.com/books?id=bDXgCO3Z4bIC"
}

@article{chiral,
    author    = "S. Frauendorf and J. Meng",
    journal = "Nucl. Phys. A",
    volume = "617",
    number = "2",
    pages = "131 - 147",
    year = "1997",
    doi = "10.1016/S0375-9474(97)00004-3"
}

@article{balabanski,
    author =       "D. L. Balabanski and M. Danchev and D. J. Hartley and L. L. Riedinger and O. Zeidan and Jing-ye Zhang and C. J. Barton and C. W. Beausang and M. A. Caprio and R. F. Casten and others",
    journal =      "Phys. Rev. C",
    volume =       "70",
    pages = 	 "044305",
    year =         "2004",
    DOI =          "10.1103/PhysRevC.70.044305"
}

@article{zhu,
  author = {Zhu, S. and Garg, U. and Nayak, B. K. and Ghugre, S. S. and Pattabiraman, N. S. and Fossan, D. B. and Koike, T. and Starosta, K. and Vaman, C. and Janssens, R. V. F. and others},
  journal = {Phys. Rev. Lett.},
  volume = {91},
  issue = {13},
  pages = {132501},
  numpages = {4},
  year = {2003},
  doi = {10.1103/PhysRevLett.91.132501}
}

@article{vaman,
  author = {Vaman, C. and Fossan, D. B. and Koike, T. and Starosta, K. and Lee, I. Y. and Macchiavelli, A. O.},
  journal = {Phys. Rev. Lett.},
  volume = {92},
  issue = {3},
  pages = {032501},
  numpages = {4},
  year = {2004},
  doi = {10.1103/PhysRevLett.92.032501}
}

@ARTICLE{wang,
  author = {S. Y. Wang and B. Qi and L. Liu and S. Q. Zhang and H. Hua and X.
	Q. Li and Y. Y. Chen and L. H. Zhu and J. Meng and S. M. Wyngaardt
	and others},
  title = {The first candidate for chiral nuclei in the mass region: $^{80}$\textsc{B}r},
  journal = {Phys. Lett. B},
  year = {2011},
  volume = {703},
  pages = {40},
  doi = "https://doi.org/10.1016/j.physletb.2011.07.055"
}

@article{odegard,
    author =       "S. W. {\O}deg{\aa}rd and G. B. Hagemann and D. R. Jensen and M. Bergstr{\"o}m and B. Herskind and G. Sletten and S. T{\"o}rm{\"a}nen and J. N. Wilson and P. O. Tj{\o}m and I. Hamamoto and others",
    journal =      "Phys. Rev. Lett.",
    volume =       "86",
    number =       "26",
    pages =        "5866--5869",
    year =         "2001",
    doi =          "10.1103/PhysRevLett.86.5866"
}

@article{Jensen,
    author =       "D. R. Jensen and G. B. Hagemann and I. Hamamoto and S. W. {\O}deg{\aa}rd and B. Herskind and G. Sletten and J. N. Wilson and K. Spohr and H. H{\"u}bel and P. Bringel and others",
    journal =      "Phys. Rev. Lett.",
    volume =       "89",
       pages =        "142503",
    year =         "2002",
    doi =          "10.1103/PhysRevLett.89.142503"
}

@article{amro,
    author =       "H. Amro and W. C. Ma and G. B. Hagemann and R. M. Diamond and J. Domscheit and P. Fallon and A. G{\"o}rgen and B. Herskind and H. H{\"u}bel and D. R. Jensen",
    journal =      "Phys. Lett. B",
    volume =       "553",
    pages =        "197--203",
    year =         "2003",
    DOI =          "10.1016/S0370-2693(02)03199-4"
}

@article{two-phonon,
  author =	"N. Sensharma and U. Garg and S. Zhu and A. D. Ayangeakaa and S. Frauendorf and W. Li and G. H. Bhat and J. A. Sheikh and M. P. Carpenter and Q. B. Chen and others",
  journal =     "Phys. Lett. B",
  volume =       "792",
  pages = 	   "170",
  year =         "2019",
  DOI =          "10.1016/j.physletb.2019.03.038"
}

@Article{Q.B.Chen2018PRC_v1,
  Title                    = {Reexamining nuclear chiral geometry from the orientation of the angular momentum},
  Author                   = {Chen, Q. B. and Meng, J.},
  Journal                  = {Phys. Rev. C},
  Year                     = {2018},
  Month                    = {Sep},
  Pages                    = {031303(R)},
  Volume                   = {98},
  Issue                    = {3},
  Numpages                 = {6},
  Publisher                = {American Physical Society}
}

@Article{Q.B.Chen2019PRC,
  Title                    = {Behavior of the collective rotor in nuclear chiral motion},
  Author                   = {Chen, Q. B. and Kaiser, N. and Mei\ss{}ner, Ulf-G. and Meng, J.},
  Journal                  = {Phys. Rev. C},
  Year                     = {2019},
  Month                    = {Jun},
  Pages                    = {064326},
  Volume                   = {99},
  DOI                      = {10.1103/PhysRevC.99.064326},
  Issue                    = {6},
  Numpages                 = {12},
  Publisher                = {American Physical Society}
}

@article{136Nd,
  title = {Experimental evidence for transverse wobbling bands in $^{136}\mathrm{Nd}$},
  author = {Lv, B. F. and Petrache, C. M. and Budaca, R. and Astier, A. and Zheng, K. K. and Greenlees, P. and Badran, H. and Calverley, T. and Cox, D. M. and Grahn, T. and others},
  journal = {Phys. Rev. C},
  volume = {105},
  issue = {3},
  pages = {034302},
  numpages = {7},
  year = {2022},
  month = {Mar},
  publisher = {American Physical Society},
  doi = {10.1103/PhysRevC.105.034302},
  url = {https://link.aps.org/doi/10.1103/PhysRevC.105.034302}
}

@article{ayangeakaa,
  title = {Evidence for Multiple Chiral Doublet Bands in $^{133}\mathrm{Ce}$},
  author = {Ayangeakaa, A. D. and Garg, U. and Anthony, M. D. and Frauendorf, S. and Matta, J. T. and Nayak, B. K. and Patel, D. and Chen, Q. B. and Zhang, S. Q. and Zhao, P. W. and others},
  journal = {Phys. Rev. Lett.},
  volume = {110},
  issue = {17},
  pages = {172504},
  numpages = {5},
  year = {2013},
  month = {Apr},
  publisher = {American Physical Society},
  doi = {10.1103/PhysRevLett.110.172504},
  url = {https://link.aps.org/doi/10.1103/PhysRevLett.110.172504}
}

@article{transverse,
    author =       "S. Frauendorf and F. D{\"o}nau",
    journal =      "Phys. Rev. C",
    volume =       "89",
    issue =       "1",
    pages = 	   "014322",
    numpages = 	   "12",
    year =         "2014",
    DOI =          "10.1103/PhysRevC.89.014322"
}

@article{radford,
    author =       "D. C. Radford",
    journal =      "Nucl. Instrum. Methods Phys. Res., Sect. A",
    volume =       "361",
    number = 	   "12",
    pages =        "297--305",
    year =         "1995",
    DOI =          "10.1016/0168-9002(95)00183-2"
}

@book{James_phd,
    author    = "J. T. Matta",
    title     = "Exotic Nuclear Excitations: The Transverse wobbling mode in $^{135}${P}r",
    year      = "2017",
    publisher = "Springer Cham",
    DOI	      = "https://doi.org/10.1007/978-3-319-53240-0",
    url     = "https://link.springer.com/book/10.1007/978-3-319-53240-0"
}

@book{springer_book,
    author    = "N. Sensharma",
    title     = "Wobbling Motion in Nuclei: Transverse, Longitudinal, and Chiral",
    year      = "2022",
    publisher = "Springer Cham",
    doi	      = "https://doi.org/10.1007/978-3-031-17150-5",
    url	      = "https://link.springer.com/book/10.1007/978-3-031-17150-5"
}

@article{au-prl,
  author = {Sensharma, N. and Garg, U. and Chen, Q. B. and Frauendorf, S. and Burdette, D. P. and Cozzi, J. L. and Howard, K. B. and Zhu, S. and Carpenter, M. P. and Copp, P. and others},
  journal = {Phys. Rev. Lett.},
  volume = {124},
  issue = {5},
  pages = {052501},
  numpages = {6},
  year = {2020},
  month = {Feb},
  publisher = {American Physical Society},
  doi = {10.1103/PhysRevLett.124.052501}
}

@Article{Q.B.Chen2018PLB,
  Title                    = {Multiple chiral doublets in four-j shells particle rotor model: Five possible chiral doublets in Nd136 },
  Author                   = {Q. B. Chen and B. F. Lv and C. M. Petrache and J. Meng},
  Journal                  = {Phys. Lett. B},
  Year                     = {2018},
  Pages                    = {744},
  Volume                   = {782},
  doi                      = {10.1016/j.physletb.2018.06.030},
}

@ARTICLE{B.Qi2009PLB,
  author = {B. Qi and S .Q. Zhang and J. Meng and S. Y. Wang and S. Frauendorf},
  title = {Chirality in odd-\textsc{$A$} nucleus $^{135}$\textsc{N}d in particle
	rotor model},
  journal = {Phys. Lett. B},
  year = {2009},
  volume = {675},
  pages = {175},
  doi   = {10.1016/j.physletb.2009.02.061}
}

@ARTICLE{J.Meng2006PRC,
  author = {Meng, J. and Peng, J. and Zhang, S. Q. and Zhou, S.-G.},
  title = {Possible existence of multiple chiral doublets in $^{106}\mathrm{Rh}$},
  journal = {Phys. Rev. C},
  year = {2006},
  volume = {73},
  pages = {037303},
  doi   = {10.1103/PhysRevC.73.037303},
}

@ARTICLE{P.W.Zhao2010PRC,
  author = {Zhao, P. W. and Li, Z. P. and Yao, J. M. and Meng, J.},
  title = {New parametrization for the nuclear covariant energy density functional
	with a point-coupling interaction},
  journal = {Phys. Rev. C},
  year = {2010},
  volume = {82},
  pages = {054319},
  month = {Nov},
  doi = {10.1103/PhysRevC.82.054319},
  issue = {5},
  numpages = {14},
  publisher = {American Physical Society}
}

@Book{J.Meng2016book,
  Title                    = {Relativistic density functional for nuclear structure},
  Editor                   = {J. Meng},
  Publisher                = {World Scientific},
  Year                     = {2016},
  Address                  = {Singapore},
  Series                   = {International Review of Nuclear Physics},
  Volume                   = {10},
  doi                      = {10.1142/9872},
}

@Article{Devi2021PLB,
  Title                    = {Observation of multiphonon transverse wobbling in 133Ba},
  Author                   = {K. {Rojeeta Devi} and Suresh Kumar and Naveen Kumar and Neelam and F. S. Babra and Md. S. R. Laskar and S. Biswas and S. Saha and P. Singh and S. Samanta and others},
  Journal                  = {Phys. Lett. B},
  Year                     = {2021},
  Pages                    = {136756},
  Volume                   = {823},
  DOI                      = {https://doi.org/10.1016/j.physletb.2021.136756}
}

@Article{Chakraborty2020PLB,
  Title                    = {Multiphonon longitudinal wobbling in 127Xe},
  Author                   = {S. Chakraborty and H. P. Sharma and S. S. Tiwary and C. Majumder and A. K. Gupta and P. Banerjee and S. Ganguly and S. Rai and Pragati and Mayank and others},
  Journal                  = {Phys. Lett. B},
  Year                     = {2020},
  Pages                    = {135854},
  Volume                   = {811},
  doi                      = {10.1016/j.physletb.2020.135854}
}

@Article{Petrache2019PLB,
  Title                    = {Diversity of shapes and rotations in the γ-soft 130Ba nucleus: First observation of a t-band in the A=130 mass region},
  Author                   = {C. M. Petrache and P. M. Walker and S. Guo and Q. B. Chen and S. Frauendorf and Y. X. Liu and R. A. Wyss and D. Mengoni and Y. H. Qiang and A. Astier and others},
  Journal                  = {Phys. Lett. B},
  Year                     = {2019},
  Pages                    = {241 - 247},
  Volume                   = {795},
  doi                      = {10.1016/j.physletb.2019.06.040},
}

@Article{Q.B.Chen2020PLB,
  Title                    = {Static quadrupole moments of nuclear chiral doublet bands},
  Author                   = {Q. B. Chen and N. Kaiser and Ulf-G. Mei{\ss}ner and J. Meng},
  Journal                  = {Phys. Lett. B},
  Year                     = {2020},
  Pages                    = {135568},
  Volume                   = {807},
  doi                      = {10.1016/j.physletb.2020.135568},
}

@ARTICLE{Starosta2001PRL,
  author = {Starosta, K. and Koike, T. and Chiara, C. J. and Fossan, D. B. and
	LaFosse, D. R. and Hecht, A. A. and Beausang, C. W. and Caprio, M.
	A. and Cooper, J. R. and Kr\"ucken, R. and others},
  title = {Chiral doublet structures in odd-odd \textit{N}$=75$ isotones: Chiral
	vibrations},
  journal = {Phys. Rev. Lett.},
  year = {2001},
  volume = {86},
  pages = {971},
  doi = {10.1103/PhysRevLett.86.971},
  issue = {6},
}

@Article{B.W.Xiong2019ADNDT,
  Title                    = {Nuclear chiral doublet bands data tables},
  Author                   = {Xiong, B. W and Wang, Y. Y.},
  Journal                  = {Atom. Data Nucl. Data Tables},
  Year                     = {2019},
  Pages                    = {193},
  Volume                   = {125},
}

@ARTICLE{Frauendorf2001RMP,
  author = {Frauendorf, S.},
  title = {Spontaneous symmetry breaking in rotating nuclei},
  journal = {Rev. Mod. Phys.},
  year = {2001},
  volume = {73},
  pages = {463},
  doi = {10.1103/RevModPhys.73.463},
  issue = {2},
}

@ARTICLE{J.Meng2010JPG,
  author = {Meng, J. and Zhang, S. Q.},
  title = {Open problems in understanding the nuclear chirality},
  journal = {J. Phys. G: Nucl. Part. Phys.},
  year = {2010},
  volume = {37},
  pages = {064025},
  doi = {10.1088/0954-3899/37/6/064025},
}

@Article{J.Meng2016PS,
  author =    {J. Meng and P. W. Zhao},
  title =     {Nuclear chiral and magnetic rotation in covariant density functional theory},
  journal =   {Phys. Scr.},
  year =      {2016},
  volume =    {91},
  pages =     {053008},
}

@article{lv,
title = {Evidence against the wobbling nature of low-spin bands in 135Pr},
journal = {Physics Letters B},
volume = {824},
pages = {136840},
year = {2022},
issn = {0370-2693},
doi = {https://doi.org/10.1016/j.physletb.2021.136840},
url = {https://www.sciencedirect.com/science/article/pii/S0370269321007802},
author = {B.F. Lv and C.M. Petrache and E.A. Lawrie and S. Guo and A. Astier and K.K. Zheng and H.J. Ong and J.G. Wang and X.H. Zhou and Z.Y. Sun and others}
}

@ARTICLE{Mukhopadhyay2007PRL,
  author = {Mukhopadhyay, S. and Almehed, D. and Garg, U. and Frauendorf, S.
	and Li, T. and Rao, P. V. Madhusudhana and Wang, X. and Ghugre, S.
	S. and Carpenter, M. P. and Gros, S. and others},
  title = {From chiral vibration to static chirality in $^{135}$\textsc{N}d},
  journal = {Phys. Rev. Lett.},
  year = {2007},
  volume = {99},
  pages = {172501}
}

@ARTICLE{Tonev2007PRC,
  author = {Tonev, D. and de Angelis, G. and Brant, S. and Frauendorf, S. and
	Petkov, P. and Dewald, A. and Doenau, F. and Balabanski, D. L. and
	Zhong, Q. and Pejovic, P. and others},
  title = {Question of dynamic chirality in nuclei: The case of $^{134}$\textsc{P}r},
  journal = {Phys. Rev. C},
  year = {2007},
  volume = {76},
  pages = {044313},
  doi = {10.1103/PhysRevC.76.044313},
}

@ARTICLE{Grodner2006PRL,
  author = {Grodner, E. and Srebrny, J. and Pasternak, A. A. and Zalewska, I.
	and Morek, T. and Droste, Ch. and Mierzejewski, J. and Kowalczyk,
	M. and Kownacki, J. and Kisielinski, M. and others},
  title = {$^{128}$\textsc{C}s as the best example revealing chiral symmetry
	breaking},
  journal = {Phys. Rev. Lett.},
  year = {2006},
  volume = {97},
  pages = {172501},
  doi = {10.1103/PhysRevLett.97.172501},
}

@ARTICLE{Tonev2006PRL,
  author = {Tonev, D. and de Angelis, G. and Petkov, P. and Dewald, A. and Brant,
	S. and Frauendorf, S. and Balabanski, D. L. and Pejovic, P. and Bazzacco,
	D. and Bednarczyk, P. and others},
  title = {Transition probabilities in $^{134}$\textsc{P}r: A test for chirality
	in nuclear systems},
  journal = {Phys. Rev. Lett.},
  year = {2006},
  volume = {96},
  pages = {052501},
  doi = {10.1103/PhysRevLett.96.052501},
}

@ARTICLE{Koike2004PRL,
  author = {Koike, T. and Starosta, K. and Hamamoto, I.},
  title = {Chiral bands, dynamical spontaneous symmetry breaking, and the selection
	rule for electromagnetic transitions in the chiral geometry},
  journal = {Phys. Rev. Lett.},
  year = {2004},
  volume = {93},
  pages = {172502},
  doi = {10.1103/PhysRevLett.93.172502},
}

@Article{Koike2003AIP,
  Title                    = {Sensitive Criterion For Chirality; Chiral Doublet Bands In 104Rh59},
  Author                   = {Koike,T. and Starosta,K. and Vaman,C. and Ahn,T. and Fossan,D. B. and Clark,R. M. and Cromaz,M. and Lee,I. Y. and Macchiavelli,A. O. },
  Journal                  = {AIP Conference Proceedings},
  Year                     = {2003},
  Number                   = {1},
  Pages                    = {160},
  Volume                   = {656},
  DOI                      = {10.1063/1.1556637},
}

@Article{Frauendorf08,
  Title                    = {Heart-shaped nuclei: Condensation of rotational-aligned octupole phonons},
  Author                   = {S. Frauendorf},
  Journal                  = {Phys. Rev. C},
  Year                     = {2008},
  Number                   = {1},
  Pages                    = {021304 (R)},
  Volume                   = {77},
  DOI                      = {10.1103/PhysRevC.77.021304},
}

@Article{Chen22,
  Title                    = {Study of wobbling modes by means of spin coherent state maps},
  Author                   = {Q. B. Chen and S. Frauendorf},
  Journal                  = {Eur. J. Phys. A},
  Year                     = {2022},
  Number                   = {1},
  Pages                    = {75},
  Volume                   = {58},
  DOI                      = {10.1140/epja/s10050-022-00727-5},
}

@ARTICLE{Zhang2007,
  author = {S. Q. Zhang and B. Qi and S. Y. Wang and J. Meng},
  title = {Chiral bands for a quasi-proton and quasi-neutron coupled with a triaxial rotor},
  journal = {Phys. Rev. C},
  year = {2007},
  volume = {75},
  pages = {044307},
  doi = {10.1103/PhysRevC.75.044307},
}

@article{163Lu,
  title = {Evidence for the Wobbling Mode in Nuclei},
  author = {\O{}deg\aa{}rd, S. W. and Hagemann, G. B. and Jensen, D. R. and Bergstr\"om, M. and Herskind, B. and Sletten, G. and T\"orm\"anen, S. and Wilson, J. N. and Tj\o{}m, P. O. and Hamamoto, I. and Spohr, K. and H\"ubel, H. and G\"orgen, A. and Sch\"onwasser, G. and Bracco, A. and Leoni, S. and Maj, A. and Petrache, C. M. and Bednarczyk, P. and Curien, D.},
  journal = {Phys. Rev. Lett.},
  volume = {86},
  issue = {26},
  pages = {5866--5869},
  numpages = {0},
  year = {2001},
  month = {Jun},
  publisher = {American Physical Society},
  doi = {10.1103/PhysRevLett.86.5866},
  url = {https://link.aps.org/doi/10.1103/PhysRevLett.86.5866}
}

@article{165Lu,
title = {One- and two-phonon wobbling excitations in triaxial $^{165}\mathrm{Lu}$},
journal = {Physics Letters B},
volume = {552},
number = {1},
pages = {9-16},
year = {2003},
issn = {0370-2693},
doi = {https://doi.org/10.1016/S0370-2693(02)03095-2},
url = {https://www.sciencedirect.com/science/article/pii/S0370269302030952},
author = {G Schönwaßer and H Hübel and G.B Hagemann and P Bednarczyk and G Benzoni and A Bracco and P Bringel and R Chapman and D Curien and J Domscheit and B Herskind and D.R Jensen and S Leoni and G {Lo Bianco} and W.C Ma and A Maj and A Neußer and S.W Ødegård and C.M Petrache and D Roßbach and H Ryde and K.H Spohr and A.K Singh}
}

@article{167Lu,
title = {The wobbling mode in $^{167}\mathrm{Lu}$},
journal = {Physics Letters B},
volume = {553},
number = {3},
pages = {197-203},
year = {2003},
issn = {0370-2693},
doi = {https://doi.org/10.1016/S0370-2693(02)03199-4},
url = {https://www.sciencedirect.com/science/article/pii/S0370269302031994},
author = {H Amro and W.C Ma and G.B Hagemann and R.M Diamond and J Domscheit and P Fallon and A Görgen and B Herskind and H Hübel and D.R Jensen and Y Li and A.O Macchiavelli and D Roux and G Sletten and J Thompson and D Ward and I Wiedenhöver and J.N Wilson and J.A Winger}
}

@article{161Lu,
title = {Evidence for wobbling excitation in $^{161}\mathrm{Lu}$},
author = {Bringel, P. and Hagemann, G. B. and Hübel, H. and Al-khatib, A. and Bednarczyk, P. and Bürger, A. and Curien, D. and Gangopadhyay, G. and Herskind, B. and Jensen, D. R. and Joss, D. T. and Kröll, Th. and Lo Bianco, G. and Lunardi, S. and Ma, W. C. and Nenoff, N. and Neußer-Neffgen, A. and Petrache, C. M. and Schönwasser, G. and Simpson, J. and Singh, A. K. and Singh, N. and Sletten, G.},
year = {2005},
journal = {The European Physical Journal A - Hadrons and Nuclei},
pages = {167 - 172},
volume = {24},
issue = {2},
url = {https://doi.org/10.1140/epja/i2005-10005-7},
doi = {10.1140/epja/i2005-10005-7}
}

@article{167Ta,
  title = {Wobbling mode in $^{167}\mathrm{Ta}$},
  author = {Hartley, D. J. and Janssens, R. V. F. and Riedinger, L. L. and Riley, M. A. and Aguilar, A. and Carpenter, M. P. and Chiara, C. J. and Chowdhury, P. and Darby, I. G. and Garg, U. and Ijaz, Q. A. and Kondev, F. G. and Lakshmi, S. and Lauritsen, T. and Ludington, A. and Ma, W. C. and McCutchan, E. A. and Mukhopadhyay, S. and Pifer, R. and Seyfried, E. P. and Stefanescu, I. and Tandel, S. K. and Tandel, U. and Vanhoy, J. R. and Wang, X. and Zhu, S. and Hamamoto, I. and Frauendorf , S.},
  journal = {Phys. Rev. C},
  volume = {80},
  issue = {4},
  pages = {041304},
  numpages = {5},
  year = {2009},
  month = {Oct},
  publisher = {American Physical Society},
  doi = {10.1103/PhysRevC.80.041304},
  url = {https://link.aps.org/doi/10.1103/PhysRevC.80.041304}
}

@article{135Pr,
  title = {Transverse Wobbling in $^{135}\mathrm{Pr}$},
  author = {Matta, J. T. and Garg, U. and Li, W. and Frauendorf, S. and Ayangeakaa, A. D. and Patel, D. and Schlax, K. W. and Palit, R. and Saha, S. and Sethi, J. and Trivedi, T. and Ghugre, S. S. and Raut, R. and Sinha, A. K. and Janssens, R. V. F. and Zhu, S. and Carpenter, M. P. and Lauritsen, T. and Seweryniak, D. and Chiara, C. J. and Kondev, F. G. and Hartley, D. J. and Petrache, C. M. and Mukhopadhyay, S. and Lakshmi, D. Vijaya and Raju, M. Kumar and Madhusudhana Rao, P. V. and Tandel, S. K. and Ray, S. and D\"onau, F.},
  journal = {Phys. Rev. Lett.},
  volume = {114},
  issue = {8},
  pages = {082501},
  numpages = {6},
  year = {2015},
  month = {Feb},
  publisher = {American Physical Society},
  doi = {10.1103/PhysRevLett.114.082501},
  url = {https://link.aps.org/doi/10.1103/PhysRevLett.114.082501}
}

@article{105Pd,
  title = {Experimental Evidence for Transverse Wobbling in $^{105}\mathrm{Pd}$},
  author = {Tim\'ar, J. and Chen, Q. B. and Kruzsicz, B. and Sohler, D. and Kuti, I. and Zhang, S. Q. and Meng, J. and Joshi, P. and Wadsworth, R. and Starosta, K. and Algora, A. and Bednarczyk, P. and Curien, D. and Dombr\'adi, Zs. and Duch\^ene, G. and Gizon, A. and Gizon, J. and Jenkins, D. G. and Koike, T. and Krasznahorkay, A. and Moln\'ar, J. and Nyak\'o, B. M. and Paul, E. S. and Rainovski, G. and Scheurer, J. N. and Simons, A. J. and Vaman, C. and Zolnai, L.},
  journal = {Phys. Rev. Lett.},
  volume = {122},
  issue = {6},
  pages = {062501},
  numpages = {6},
  year = {2019},
  month = {Feb},
  publisher = {American Physical Society},
  doi = {10.1103/PhysRevLett.122.062501},
  url = {https://link.aps.org/doi/10.1103/PhysRevLett.122.062501}
}

@article{nandi,
  title = {First Observation of Multiple Transverse Wobbling Bands of Different Kinds in $^{183}\mathrm{Au}$},
  author = {Nandi, S. and Mukherjee, G. and Chen, Q. B. and Frauendorf, S. and Banik, R. and Bhattacharya, Soumik and Dar, Shabir and Bhattacharyya, S. and Bhattacharya, C. and Chatterjee, S. and Das, S. and Samanta, S. and Raut, R. and Ghugre, S. S. and Rajbanshi, S. and Ali, Sajad and Pai, H. and Asgar, Md. A. and Das Gupta, S. and Chowdhury, P. and Goswami, A.},
  journal = {Phys. Rev. Lett.},
  volume = {125},
  issue = {13},
  pages = {132501},
  numpages = {6},
  year = {2020},
  month = {Sep},
  publisher = {American Physical Society},
  doi = {10.1103/PhysRevLett.125.132501},
  url = {https://link.aps.org/doi/10.1103/PhysRevLett.125.132501}
}

@article{187Au,
  title = {Longitudinal Wobbling Motion in $^{187}\mathrm{Au}$},
  author = {Sensharma, N. and Garg, U. and Chen, Q. B. and Frauendorf, S. and Burdette, D. P. and Cozzi, J. L. and Howard, K. B. and Zhu, S. and Carpenter, M. P. and Copp, P. and Kondev, F. G. and Lauritsen, T. and Li, J. and Seweryniak, D. and Wu, J. and Ayangeakaa, A. D. and Hartley, D. J. and Janssens, R. V. F. and Forney, A. M. and Walters, W. B. and Ghugre, S. S. and Palit, R.},
  journal = {Phys. Rev. Lett.},
  volume = {124},
  issue = {5},
  pages = {052501},
  numpages = {6},
  year = {2020},
  month = {Feb},
  publisher = {American Physical Society},
  doi = {10.1103/PhysRevLett.124.052501},
  url = {https://link.aps.org/doi/10.1103/PhysRevLett.124.052501}
}

@article{130Ba,
  title = {Transverse wobbling in an even-even nucleus},
  author = {Chen, Q. B. and Frauendorf, S. and Petrache, C. M.},
  journal = {Phys. Rev. C},
  volume = {100},
  issue = {6},
  pages = {061301},
  numpages = {5},
  year = {2019},
  month = {Dec},
  publisher = {American Physical Society},
  doi = {10.1103/PhysRevC.100.061301},
  url = {https://link.aps.org/doi/10.1103/PhysRevC.100.061301}
}

@article{tw_frauendorf,
  title = {Transverse wobbling: A collective mode in odd-$A$ triaxial nuclei},
  author = {Frauendorf, S. and D\"onau, F.},
  journal = {Phys. Rev. C},
  volume = {89},
  issue = {1},
  pages = {014322},
  numpages = {12},
  year = {2014},
  month = {Jan},
  publisher = {American Physical Society},
  doi = {10.1103/PhysRevC.89.014322},
  url = {https://link.aps.org/doi/10.1103/PhysRevC.89.014322}
}

@article{135Pr_yrast,
  title = {High-spin yrast states in the $\ensuremath{\gamma}$-soft nuclei ${}^{135}$Pr and ${}^{134}$Ce},
  author = {Paul, E. S. and Fox, C. and Boston, A. J. and Chantler, H. J. and Chiara, C. J. and Clark, R. M. and Cromaz, M. and Descovich, M. and Fallon, P. and Fossan, D. B. and Hecht, A. A. and Koike, T. and Lee, I. Y. and Macchiavelli, A. O. and Nolan, P. J. and Starosta, K. and Wadsworth, R. and Ragnarsson, I.},
  journal = {Phys. Rev. C},
  volume = {84},
  issue = {4},
  pages = {047302},
  numpages = {4},
  year = {2011},
  month = {Oct},
  publisher = {American Physical Society},
  doi = {10.1103/PhysRevC.84.047302},
  url = {https://link.aps.org/doi/10.1103/PhysRevC.84.047302}
}

@misc{MCMCMethod,
      title={Structure effects on the giant monopole resonance and determinations of the nuclear incompressibility}, 
      author={K. B. Howard},
      year={2020},
      eprint={2004.02362},
      archivePrefix={arXiv},
      primaryClass={nucl-ex},
      url={https://arxiv.org/abs/2004.02362}, 
}

@misc{lawrie,
  author = "E. Lawrie",
  howpublished = "private communication"
}

@article{timar2011,
  title = {Medium- and high-spin band structure of the chiral-candidate nucleus ${}^{134}$Pr},
  author = {Tim\'ar, J. and Starosta, K. and Kuti, I. and Sohler, D. and Fossan, D. B. and Koike, T. and Paul, E. S. and Boston, A. J. and Chantler, H. J. and Descovich, M. and Clark, R. M. and Cromaz, M. and Fallon, P. and Lee, I. Y. and Macchiavelli, A. O. and Chiara, C. J. and Wadsworth, R. and Hecht, A. A. and Almehed, D. and Frauendorf, S.},
  journal = {Phys. Rev. C},
  volume = {84},
  issue = {4},
  pages = {044302},
  numpages = {16},
  year = {2011},
  month = {Oct},
  publisher = {American Physical Society},
  doi = {10.1103/PhysRevC.84.044302},
  url = {https://link.aps.org/doi/10.1103/PhysRevC.84.044302}
}

@Article{CF2025,
  Title                    = {Entanglement in two-quasiparticle-triaxial-rotor systems: Chirality, wobbling, and the Pauli effect},
  Author                   = {Chen, Q. B. and Frauendorf, S.},
  Journal                  = {Phys. Rev. C},
  Year                     = {2025},
  Month                    = {Jun},
  Pages                    = {064326},
  Volume                   = {111},
  DOI                      = {10.1103/s6zp-h83d},
  Issue                    = {6},
  Numpages                 = {17},
  Publisher                = {American Physical Society}
}

@Article{Rajbanshi2025PRC,
  Title                    = {Conclusive evidence of a two-neutron multiphonon transverse wobbling mode in $^{82}\mathrm{Kr}$},
  Author                   = {Rajbanshi, S. and Palit, R. and Rahaman, Habibur and Manna, G. and Ali, Sajad and Chakraborty, S. and Bhat, G. H. and Jehangir, S. and Sheikh, J. A. and Babra, F. S. and Banik, R. and Bhattacharya, S. and Bhattacharyya, S. and Dey, P. and Laskar, Md. S. R. and Mukherjee, G. and Nandi, S. and Pai, H. and Santra, Rajkumar and Trivedi, T.},
  Journal                  = {Phys. Rev. C},
  Year                     = {2025},
  Month                    = {Jun},
  Pages                    = {L061301},
  Volume                   = {111},
  DOI                      = {10.1103/PhysRevC.111.L061301},
  Issue                    = {6},
  Numpages                 = {7},
  Publisher                = {American Physical Society}
}

@Article{Biswas2019EPJA,
  Title                    = {Longitudinal wobbling in $^{133}\textrm{La}$},
  Author                   = {Biswas, S. and Palit, R. and Frauendorf, S. and Garg, U. and Li, W. and Bhat, G. H. and Sheikh, J. A. and Sethi, J. and Saha, S. and Singh, Purnima and Choudhury, D. and Matta, J. T. and Ayangeakaa, A. D. and Dar, W. A. and Singh, V. and Sihotra, S.},
  Journal                  = {Eur. Phys. J. A},
  Year                     = {2019},
  Pages                    = {159},
  Volume                   = {55},
  DOI                      = {10.1140/epja/i2019-12856-5}
}

@Article{Prajapati2024PRC,
  Title                    = {Possible wobbling phenomenon in $^{125}\mathrm{Xe}$},
  Author                   = {Prajapati, Mamta and Nag, Somnath and Singh, A. K. and Al-Khatib, A. and H\"ubel, H. and Neu\ss{}er-Neffgen, A. and Hagemann, G. B. and Sletten, G. and Herskind, B. and Hansen, C. R. and Benzoni, G. and Bracco, A. and Janssens, R. V. F. and Carpenter, M. P. and Chowdhury, P.},
  Journal                  = {Phys. Rev. C},
  Year                     = {2024},
  Month                    = {Mar},
  Pages                    = {034301},
  Volume                   = {109},
  DOI                      = {10.1103/PhysRevC.109.034301},
  Issue                    = {3},
  Numpages                 = {7},
  Publisher                = {American Physical Society}
}

@Article{Chakraborty2024PRC,
  Title                    = {Exploring the possibility of wobbling motion in $^{129}\mathrm{Ba}$},
  Author                   = {Chakraborty, S. and Bhattacharyya, S. and Mukherjee, G. and Majumder, C.},
  Journal                  = {Phys. Rev. C},
  Year                     = {2024},
  Month                    = {Aug},
  Pages                    = {024324},
  Volume                   = {110},
  DOI                      = {10.1103/PhysRevC.110.024324},
  Issue                    = {2},
  Numpages                 = {5},
  Publisher                = {American Physical Society}
}

@Article{Rajbanshi2024PRC,
  Title                    = {Signatures of transverse wobbling motion in $^{139}\mathrm{Pm}$},
  Author                   = {Rajbanshi, S. and Manna, G. and Rahaman, Habibur and Ali, Sajad and Nag, Somnath and Palit, R. and Pai, H. and Chakraborty, S. and Bhattacharyya, S. and Mukherjee, G. and Saha, S. and Sethi, J. and Singh, A. K. and Trivedi, T.},
  Journal                  = {Phys. Rev. C},
  Year                     = {2024},
  Month                    = {Oct},
  Pages                    = {044315},
  Volume                   = {110},
  DOI                      = {10.1103/PhysRevC.110.044315},
  Issue                    = {4},
  Numpages                 = {15},
  Publisher                = {American Physical Society}
}

@Article{Mukherjee2023PRC,
  Title                    = {Evidence of transverse wobbling motion in $^{151}\mathrm{Eu}$},
  Author                   = {Mukherjee, A. and Bhattacharya, S. and Trivedi, T. and Tiwari, S. and Singh, R. P. and Muralithar, S. and Yashraj and Katre, K. and Kumar, R. and Palit, R. and Chakraborty, S. and Jehangir, S. and Nazir, Nazira and Rouoof, S. P. and Bhat, G. H. and Sheikh, J. A. and Rather, N. and Raut, R. and Ghugre, S. S. and Ali, S. and Rajbanshi, S. and Nag, S. and Tiwary, S. S. and Sharma, A. and Kumar, S. and Yadav, S. and Jain, A. K.},
  Journal                  = {Phys. Rev. C},
  Year                     = {2023},
  Month                    = {May},
  Pages                    = {054310},
  Volume                   = {107},
  DOI                      = {10.1103/PhysRevC.107.054310},
  Issue                    = {5},
  Numpages                 = {11},
  Publisher                = {American Physical Society}
}

@Article{Hamilton2010NPA,
  Title                    = {Super deformation to maximum triaxiality in A=100–112; superdeformation, chiral bands and wobbling motion},
  Author                   = {J. H. Hamilton and S. J. Zhu and Y. X. Luo and A. V. Ramayya and S. Frauendorf and J. O. Rasmussen and J. K. Hwang and S. H. Liu and G. M. Ter-Akopian and A. V. Daniel and Y. Oganessian},
  Journal                  = {Nucl. Phys. A},
  Year                     = {2010},
  Number                   = {1},
  Pages                    = {28c - 31c},
  Volume                   = {834},
  DOI                      = {https://doi.org/10.1016/j.nuclphysa.2010.01.010}
}

@Conference{Y.X.Luo2013proceeding,
  Title                    = {Triaxial and triaxial softness in neutron rich \textsc{R}u and \textsc{P}d nuclei},
  Author                   = {Y. X. Luo and J. H. Hamilton and A. V. Ramayya and J. K. Hwang and S. H. Liu and J. O. Rasmussen and S. Frauendorf and G. M. Ter-Akopian and A. V. Daniel and Yu Ts Oganessian},
  BookTitle                = {Exotic nuclei: Exon-2012: Proceedings of the international symposium},
  Year                     = {2013},
  DOI                      = {https://doi.org/10.1142/9789814508865_0030}
}

@InBook{Frauendorf2024book,
  Title                    = {Chirality and wobbling in atomic nuclei},
  Author                   = {S. Frauendorf},
  Chapter                  = {2},
  Editor                   = {C. M. Petrache},
  Pages                    = {24},
  Publisher                = {CRC Press},
  Year                     = {2024},
  DOI                      = {https://doi.org/10.1201/9781032691633}
}

@Article{R.J.Guo2024PRL,
  Title                    = {Evidence for Chiral Wobbler in Nuclei},
  Author                   = {Guo, R. J. and Wang, S. Y. and Liu, C. and Bark, R. A. and Meng, J. and Zhang, S. Q. and Qi, B. and Rohilla, A. and Li, Z. H. and Hua, H. and Chen, Q. B. and Jia, H. and Lu, X. and Wang, S. and Sun, D. P. and Han, X. C. and Xu, W. Z. and Wang, E. H. and Bai, H. F. and Li, M. and Jones, P. and Sharpey-Schafer, J. F. and Wiedeking, M. and Shirinda, O. and Brits, C. P. and Malatji, K. L. and Dinoko, T. and Ndayishimye, J. and Mthembu, S. and Jongile, S. and Sowazi, K. and Kutlwano, S. and Bucher, T. D. and Roux, D. G. and Netshiya, A. A. and Mdletshe, L. and Noncolela, S. and Mtshali, W.},
  Journal                  = {Phys. Rev. Lett.},
  Year                     = {2024},
  Month                    = {Feb},
  Pages                    = {092501},
  Volume                   = {132},
  DOI                      = {10.1103/PhysRevLett.132.092501},
  Issue                    = {9},
  Numpages                 = {7},
  Publisher                = {American Physical Society}
}

@ARTICLE{Kuti2014PRL,
  author = {Kuti, I. and Chen, Q. B. and Tim\'ar, J. and Sohler, D. and Zhang,
	S. Q. and Zhang, Z. H. and Zhao, P. W. and Meng, J. and Starosta,
	K. and Koike, T. and Paul, E. S. and Fossan, D. B. and Vaman, C.},
  title = {Multiple chiral doublet bands of identical configuration in $^{103}\mathrm{Rh}$},
  journal = {Phys. Rev. Lett.},
  year = {2014},
  volume = {113},
  pages = {032501},
  doi = {10.1103/PhysRevLett.113.032501},
  issue = {3},
  numpages = {5},
  publisher = {American Physical Society}
}

@Article{Krako2024PLB,
  Title                    = {Multiple chiral doublet bands in $^{104}$Rh},
  Author                   = {A. Krak\'{o} and D. Sohler and J. Tim\'{a}r and I. Kuti and Q. B. Chen and S. Q. Zhang and J. Meng and K. Starosta and T. Koike and E.S. Paul and D. B. Fossan and C. Vaman},
  Journal                  = {Phys. Lett. B},
  Year                     = {2024},
  Pages                    = {138850},
  Volume                   = {855},
  DOI                      = {https://doi.org/10.1016/j.physletb.2024.138850},
}

@Article{C.Liu2016PRL,
  Title                    = {Evidence for Octupole Correlations in Multiple Chiral Doublet Bands},
  Author                   = {Liu, C. and Wang, S. Y. and Bark, R. A. and Zhang, S. Q. and Meng, J. and Qi, B. and Jones, P. and Wyngaardt, S. M. and Zhao, J. and Xu, C. and Zhou, S.-G. and Wang, S. and Sun, D. P. and Liu, L. and Li, Z. Q. and Zhang, N. B. and Jia, H. and Li, X. Q. and Hua, H. and Chen, Q. B. and Xiao, Z. G. and Li, H. J. and Zhu, L. H. and Bucher, T. D. and Dinoko, T. and Easton, J. and Juh\'asz, K. and Kamblawe, A. and Khaleel, E. and Khumalo, N. and Lawrie, E. A. and Lawrie, J. J. and Majola, S. N. T. and Mullins, S. M. and Murray, S. and Ndayishimye, J. and Negi, D. and Noncolela, S. P. and Ntshangase, S. S. and Nyak\'o, B. M. and Orce, J. N. and Papka, P. and Sharpey-Schafer, J. F. and Shirinda, O. and Sithole, P. and Stankiewicz, M. A. and Wiedeking, M.},
  Journal                  = {Phys. Rev. Lett.},
  Year                     = {2016},
  Month                    = {Mar},
  Pages                    = {112501},
  Volume                   = {116},
  DOI                      = {10.1103/PhysRevLett.116.112501},
  Issue                    = {11},
  Numpages                 = {6},
  Publisher                = {American Physical Society}
}

@Article{Grodner2018PRL,
  Title                    = {First Measurement of the $g$ Factor in the Chiral Band: The Case of the $^{128}\mathrm{Cs}$ Isomeric State},
  Author                   = {Grodner, E. and Srebrny, J. and Droste, Ch. and Pr\'ochniak, L. and Rohozi\ifmmode \acute{n}\else \'{n}\fi{}ski, S. G. and Kowalczyk, M. and Ionescu-Bujor, M. and Ur, C. A. and Starosta, K. and Ahn, T. and Kisieli\ifmmode \acute{n}\else \'{n}\fi{}ski, M. and Marchlewski, T. and Aydin, S. and Recchia, F. and Georgiev, G. and Lozeva, R. and Fiori, E. and Zieli\ifmmode \acute{n}\else \'{n}\fi{}ska, M. and Chen, Q. B. and Zhang, S. Q. and Yu, L. F. and Zhao, P. W. and Meng, J.},
  Journal                  = {Phys. Rev. Lett.},
  Year                     = {2018},
  Month                    = {Jan},
  Pages                    = {022502},
  Volume                   = {120},
  DOI                      = {10.1103/PhysRevLett.120.022502},
  Issue                    = {2},
  Numpages                 = {5},
  Publisher                = {American Physical Society}
}

@ARTICLE{Tonev2014PRL,
  author = {Tonev, D. and Yavahchova, M. S. and Goutev, N. and de Angelis, G.
	and Petkov, P. and Bhowmik, R. K. and Singh, R. P. and Muralithar,
	S. and Madhavan, N. and Kumar, R. and Kumar Raju, M. and Kaur, J.
	and Mohanto, G. and Singh, A. and Kaur, N. and Garg, R. and Shukla,
	A. and Marinov, Ts. K. and Brant, S.},
  title = {Candidates for twin chiral bands in $^{102}\mathrm{Rh}$},
  journal = {Phys. Rev. Lett.},
  year = {2014},
  volume = {112},
  pages = {052501},
  doi = {10.1103/PhysRevLett.112.052501},
  issue = {5},
  numpages = {5},
  publisher = {American Physical Society}
}

\end{document}